\documentclass[twocolumn,apj,numberedappendix]{emulateapj}
\usepackage{ctable}
\usepackage{amsmath}
\usepackage{graphicx}
\usepackage{hyperref}
\usepackage{mathrsfs}
\usepackage{natbib}

\def\b{\mathbf{b}}

\newcommand{\x}{\mathbf{x}}
\newcommand{\xhat}{\hat{\mathbf{x}}}

\newcommand{\y}{\mathbf{y}}
\newcommand{\N}{\mathbf{N}}
\newcommand{\Hmat}{\mathbf{H}}
\newcommand{\Nfg}{\mathbf{N}_{\textrm{fg}}}
\newcommand{\Q}{\mathbf{Q}}
\newcommand{\M}{\mathbf{M}}
\newcommand{\W}{\mathbf{W}}

\newcommand{\R}{\mathbf{R}}
\newcommand{\F}{\mathbf{F}}
\newcommand{\rhat}{\hat{\mathbf{r}}}
\newcommand{\Nbl}{N_{\textrm{bl}}}

\newcommand{\acl}[1]{}
\newcommand{\mep}[1]{}

\begin{document}

\title{Measuring the cosmological $21\,\textrm{cm}$ monopole with an interferometer}

\author{Morgan E. Presley\altaffilmark{1},
Adrian Liu\altaffilmark{2,3},
Aaron R. Parsons\altaffilmark{2,4}
}
\altaffiltext{1}{Department of Astrophysical Sciences, Princeton University, Princeton, NJ 08544, USA}
\altaffiltext{2}{Department of Astronomy, UC Berkeley, Berkeley, CA 94720, USA}
\altaffiltext{3}{Berkeley Center for Cosmological Physics, UC Berkeley, Berkeley, CA 94720, USA}
\altaffiltext{4}{Radio Astronomy Laboratory, UC Berkeley, Berkeley, CA 94720, USA}
\email{acliu@berkeley.edu}

\begin{abstract}
A measurement of the cosmological $21\,\textrm{cm}$ signal remains a promising but as-of-yet unattained ambition of radio astronomy. A positive detection would provide direct observations of key unexplored epochs of our cosmic history, including the cosmic dark ages and reionization. In this paper, we concentrate on measurements of the spatial monopole of the $21\,\textrm{cm}$ brightness temperature as a function of redshift (the ``global signal"). Most global experiments to date have been single-element experiments. In this paper, we show how an interferometer can be designed to be sensitive to the monopole mode of the sky, thus providing an alternate approach to accessing the global signature. We provide simple rules of thumb for designing a global signal interferometer and use numerical simulations to show that a modest array of tightly packed antenna elements with moderately sized primary beams (full-width-half-max of $\sim 40^\circ$) can compete with typical single-element experiments in their ability to constrain phenomenological parameters pertaining to reionization and the pre-reionization era. We also provide a general data analysis framework for extracting the global signal from interferometric measurements (with analysis of single-element experiments arising as a special case) and discuss trade-offs with various data analysis choices. Given that interferometric measurements are able to avoid a number of systematics inherent in single-element experiments, our results suggest that interferometry ought to be explored as a complementary way to probe the global signal.
%

\end{abstract}


\section{Introduction}

While recent years have marked tremendous progress in astronomical measurements at increasingly high redshifts, still missing are direct observations of our Universe when the first generation of luminous objects were being formed. Such observations would provide constraints on crucial periods in our cosmic timeline, including the epoch of reionization, when the intergalactic medium (IGM) experienced a large-scale phase transition, changing from neutral to almost fully ionized. Optical and infrared observations at $z \lesssim 7$ have provided some constraints on the end stages of reionization \citep{fan_et_al2006,bolton_et_al2011,treu_et_al2013,Faisst_et_al2014}, but have difficulties probing its early to intermediate stages. Moreover, modeling uncertainties often make observations difficult to interpret \citep{Dijkstra2014,Lidz2014}. Cosmic microwave background (CMB) experiments are sensitive to secondary anisotropies sourced by reionization \citep{WMAP9,zahn_et_al2012,George2014}, but the resulting constraints are based on measurements integrated along one's line-of-sight and are at best rather coarse probes of the relevant astrophysics. These existing probes have even more difficulty pushing beyond reionization to the earlier epoch known as the dark ages, during which time the first stars were formed.

One promising way to directly observe both reionization and the dark ages would be to make use of the $21\,\textrm{cm}$ hyperfine transition of hydrogen \citep{Madau_etal_1997}. By statistically measuring the brightness temperature of the $21\,\textrm{cm}$ line, one probes both the distribution of large scale structure (using atomic hydrogen as a tracer) and the ionization state of the IGM. Given the abundance of neutral hydrogen at a broad range of redshifts through the end of reionization, the $21\,\textrm{cm}$ line is an ideal way to place direct constraints on the first luminous objects and how they affected their surroundings (see, e.g., \citealt{Furlanetto2006,Morales2010,Pritchard2012,AviBook} for reviews). Moreover, the spectral nature of $21\,\textrm{cm}$ measurements not only allows a three-dimensional reconstruction of the brightness temperature distribution, but also provides information on its evolution.

\begin{figure*}[!]
	\centering
	\includegraphics[width=1.00\textwidth] {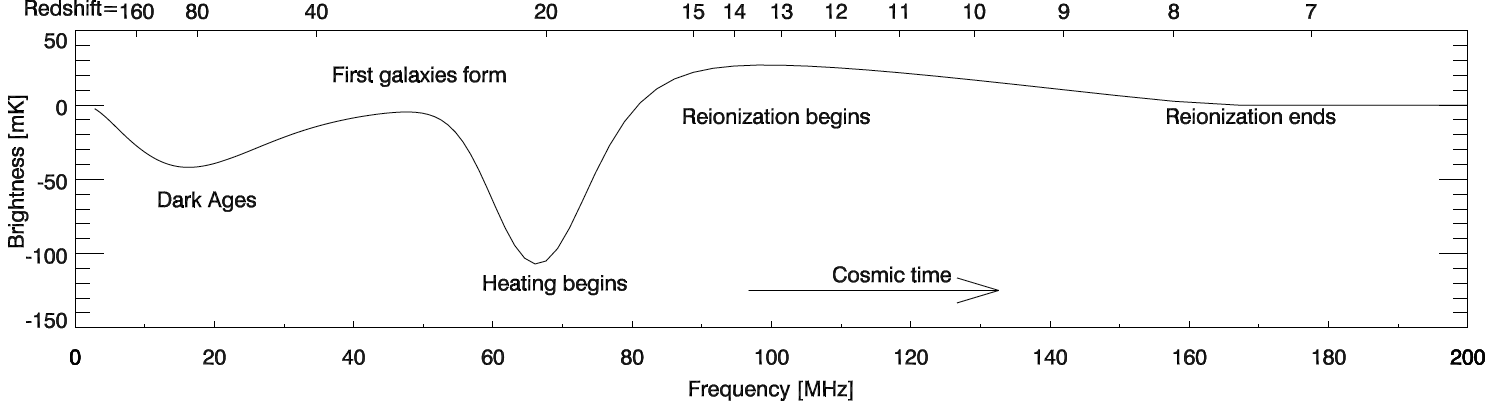}
	\caption{A fiducial model of the global 21 cm signal, as modeled in \citet{PritchardLoeb2010}. Although this model should capture the essential features of the signal, the precise details have yet to be confirmed and depend on the nature of the first stars and galaxies.}
	\label{fig:21cmSignal}
\end{figure*}

Just as with the CMB, the $21\,\textrm{cm}$ line can be characterized by a mean brightness temperature (obtained by averaging the cosmological signal in angle over the entire sky) and anisotropic fluctuations about this mean. However, unlike the CMB, the mean $21\,\textrm{cm}$ brightness temperature does not follow a simple blackbody spectrum. Instead, this ``global $21\,\textrm{cm}$ signal" is richly dependent on the astrophysics of the dark ages and reionization \citep{Shaver1999,PritchardLoeb2010,MorandiBarkana}. Figure \ref{fig:21cmSignal} shows a schematic of a fiducial model of the global 21 cm signal, highlighting the important epochs and corresponding features in the signal. The first epoch, the cosmic ``dark ages,'' arises with the thermal decoupling of the 21 cm spin states from the cosmic microwave background (CMB) and is marked by a shallow absorption feature. As the gas density continues to fall with the universe's expansion, collisions are no longer able to couple the spin states to the gas, and the signal falls back into coupling with the CMB. The next epoch is marked by the formation of the first stars and galaxies, whose Ly$\alpha$ photons strongly couple the spin states to the gas temperature. This first results in a deep absorption feature, as the gas temperature is far below that of the CMB. (Although see \citealt{GnedinShaver2004} for an example where shock heating can reduce the depth of the feature). Eventually, heating from X-ray emission pushes the gas above the CMB temperature, resulting in a 21 cm emission signal. This leads to the final epoch, the ``epoch of reionization,'' where UV photons ionize the gas, gradually erasing the 21 cm signal.

A measurement of the 21 cm global signal would also have the potential to rule out other models such as those involving dark matter annihilations or stellar black holes. Dark matter annihilation scenarios provide heating beyond that from X-ray emission and hence dampen the absorption and emission signals at the end of the dark ages and during the epoch of reionization \citep{Valdes2013_DM,Evoli2014}. Similarly, ionizing photons from accreting stellar black holes might also add significant heating. Work by \citet{Mirabel_stellar_bh} suggests that including the effects of black holes would cause the 21 cm emission signal to occur earlier than otherwise expected and would also shorten the width of the absorption feature. A global 21 cm measurement would provide evidence for or against such models.

Unfortunately, the low-frequency observations required for a measurement of the global $21\,\textrm{cm}$ signal are technically challenging. For example, the lowest frequencies are strongly affected by ionospheric fluctuations \citep{VedanthamIonosphere,DattaIonosphere,Rogers_etal_Ionosphere}, which imprint systematics in the final measured spectra. For this reason, the 
highest redshift dip (labeled ``Dark Ages" in Figure \ref{fig:21cmSignal}) will be extremely difficult to measure, and in this paper we concentrate on forecasts for observations targeting the absorption feature at $\sim 70 \,\textrm{MHz}$ (henceforth denoted the ``pre-reionization dip") and the gradual decay of the signal due to reionization from $100$ to $200\,\textrm{MHz}$.

\begin{figure}[!]
	\centering
	\includegraphics[width=0.5\textwidth] {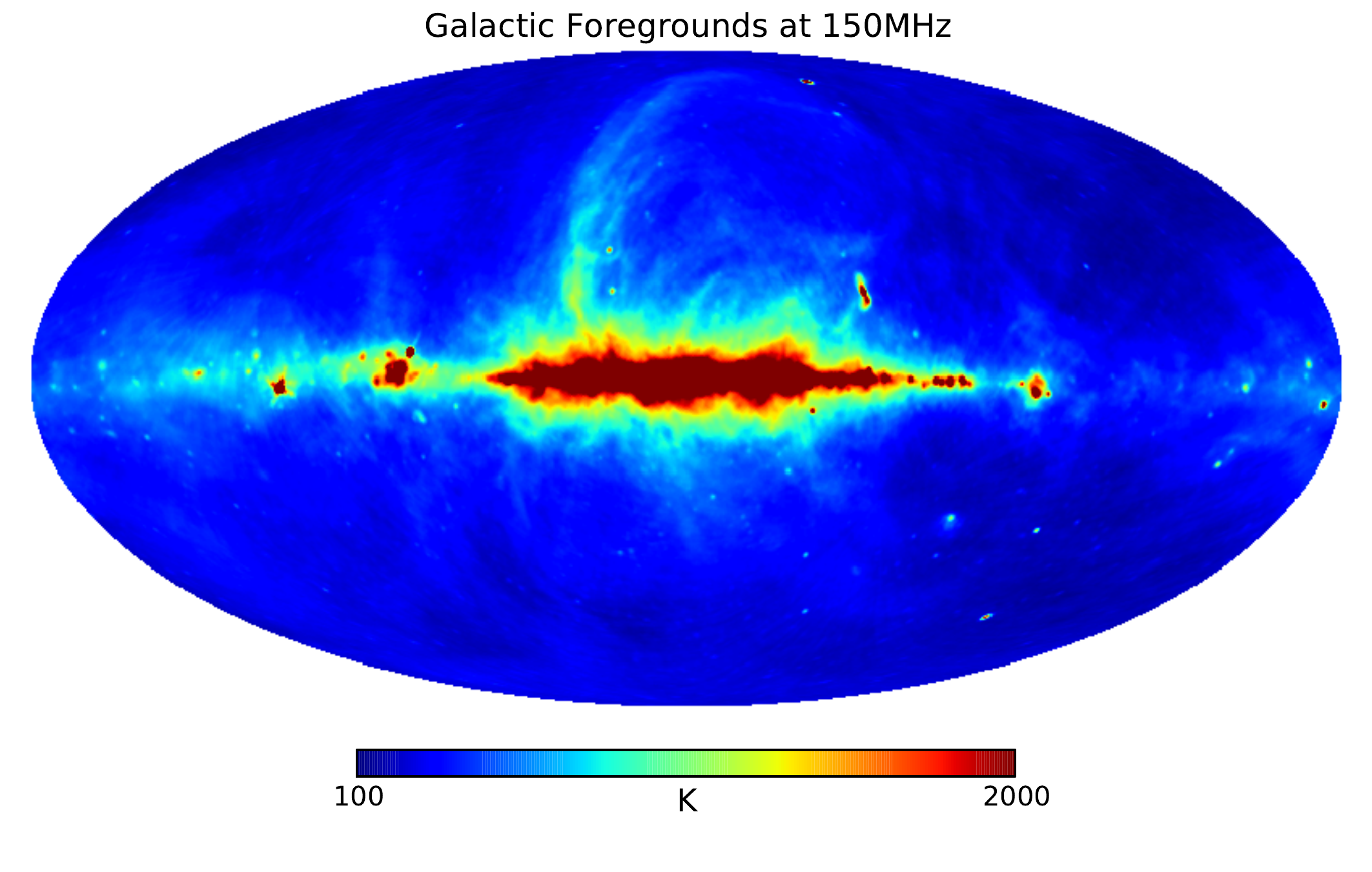}
	\caption{An empirically motivated model of Galactic synchrotron emission from \citet{angelicaGSM}. Foregrounds such as Galactic synchrotron radiation dominate the cosmological signal and must be removed from the data.}
	\label{fig:GSM}
\end{figure}

In addition, foreground contamination is a serious concern. Consider Figure \ref{fig:GSM}, for example, where we show a model of Galactic synchrotron radiation at $150\,\textrm{MHz}$ from \citet{angelicaGSM}. Even in the coolest parts of the sky (e.g., the Galactic poles), the synchrotron foregrounds dwarf the cosmological signal, as we can see from examining the scales on Figure \ref{fig:21cmSignal}. Moreover, foregrounds get brighter as one moves to lower and lower frequencies, which again makes observations increasingly challenging as one moves to higher and higher redshifts. The same is true for other foreground sources, such as extragalactic point sources, whether resolved or part of an unresolved background.

To access the cosmological signal, one must subtract or fit out the foregrounds. Some proposals (e.g., \citealt{L13}) take advantage of the angular structure of foregrounds to aid foreground mitigation, with the reasoning that any non-monopole signature on the sky cannot be the cosmological global signal. However, since the foregrounds do contain a monopole mode, there must ultimately be some subtraction of foregrounds in the final spectra. Most data analysis pipelines assume that foregrounds are spectrally smooth, and access the cosmological signal by fitting smooth functional forms to the initial spectra. Key to the success of this strategy is the assumption that the foregrounds can be modeled with a relatively small number of parameters. Otherwise, there is the danger of overfitting the spectrum, which destroys cosmological information. For this assumption to hold true, instrumental systematics must be exquisitely controlled. An unsmooth spectral ripple in the instrument, for example, will imprint chromatic signatures in the measured foregrounds, increasing the number of fitting parameters needed for their removal. Although there may be ways to mitigate these systematics in data analysis \citep{Liu_Switzer_2014}, it is best to not incur them in the first place.

There are currently a large number of experiments seeking to make a first detection and characterization of the global $21\,\textrm{cm}$ signal.  The Experiment to Detect the Global EoR Signal (EDGES) uses an extremely well-calibrated single element \citep{rogersCalib} to integrate over large parts of the sky, producing a global spectrum from $100$ to $200\,\textrm{MHz}$.  Modeling foreground spectra as a sum of low-order polynomials, EDGES has placed a lower limit of $\Delta z > 0.06$ on the duration of reionization \citep{bowmanRogersMeasurement}.  Similar in concept but operating at a lower frequency range of $55$ to $99\,\textrm{MHz}$ is the Sonda Cosmol\'{o}gica de las Islas para la Detecci\'{o}n de Hidr\'{o}geno Neutro (SCI-HI) experiment.  This frequency range corresponds to the redshift range $13.3 < z < 24.9$, providing access to the prominent dip in the signal prior to reionization.  Using a similar polynomial foreground subtraction technique to EDGES, SCI-HI is able to achieve a foreground residual level of $\sim 10\,\textrm{K}$ at $\sim 70 \,\textrm{MHz}$ \citep{voytekSCIHI}. Other single-element systems include the Shaped Antenna measurement of the background RAdio Spectrum (SARAS; \citealt{PatraSARAS}) and Broadband Instrument for Global HydrOgen ReioNisation Signal (BIGHORNS; \citealt{BIGHORNS}).

To escape radio frequency interference (RFI), most of these experiments are deployed in remote locations. For example, EDGES observes from the Murchison Radio-astronomy Observatory in Western Australia, while SCI-HI has been deployed at Isla Guadalupe in Mexico, with plans to observe at Isla Socorro and/or Isla Clari\'{o}n in the future.  To achieve even better RFI isolation, as well as to escape ionospheric distortions, the Dark Ages Radio Explorer (DARE) satellite has been proposed \citep{BurnsDARE,DAREMCMC}.  DARE consists of a short dipole antenna in lunar orbit, which allows the Moon to be used as an RFI shield.  DARE probes a frequency range of $40$ to $120\,\textrm{MHz}$, again providing direct access to the pre-reionization epoch.

Moving beyond single element experiments, \citet{Mahesh_et_al2014} and \citet{Singh_et_al2015} have explored the possibility of extracting an auto-correlation from a cross-correlation of two elements by using a resistive fence to act as a radio-wavelength analog to a beam-splitter. Placing the fence between the two elements of a baseline effectively creates a zero-baseline interferometer, which is sensitive to the monopole. Continuing to increase the number of elements, the Large-aperture Experiment to detect the Dark Ages (LEDA) makes use of a full interferometric array of antennas to simultaneously model the sky and calibration parameters \citep{BernardiLEDA}.  Fundamentally, however, its measurement of the global signal is still expected to come from total power measurements (i.e., auto-correlations) from single elements treated independently.  This differs from the approach taken by \citet{McKinley_et_al2013} and \citet{VedanthamLOFAR2}, where the LOw Frequency ARray (LOFAR) was operated as a true interferometer not just for calibration purposes, but also for the cosmological measurement itself.  At a basic level, one might imagine that interferometers are sensitive only to spatially fluctuating signals on the sky (if one follows the standard procedure of avoiding noise bias by discarding auto-correlations in the data), and are therefore insensitive to the global signal.  However, an externally imposed spatial dependence can introduce sufficient spatial structure for a global signal to be measurable by an interferometer.  \citet{VedanthamLOFAR2} took advantage of this fact by observing fields containing the Moon, effectively using lunar occultations to introduce the necessary spatial dependence for an interferometric measurement of the global signal.  So far, this approach has yielded a reasonably high signal-to-noise characterization of the foreground contaminants between $\nu =35$ and $80\,\textrm{MHz}$.

While perhaps slightly more complicated to construct than single-element experiments, interferometers can potentially provide easier control of certain instrumental systematics. For example, by omitting the auto-correlation mode of an antenna with itself, an interferometric experiment avoids the systematic noise bias that would have to be modeled and subtracted off in a single-element experiment. This noise, which arises in amplification stages, typically has significant spectral structure that can be crippling for a global $21\,\textrm{cm}$ experiment. Flux scale calibration may also be easier with multiple elements, since the elements can be coherently phased to bright astronomical sources with known positions. In this paper, we build on \citet{VedanthamLOFAR2}, generalizing their work to consider a general theory of interferometric global signal measurements.  We provide a mathematically rigorous framework for extracting the global signal from an interferometer. We also provide guiding principles for the design of a global signal interferometer. Performing numerical simulations of a fiducial interferometer, we find that interferometry can be a competitive way to probe the global signal.

The rest of this paper is organized as follows.  In Section \ref{sec:BackOfEnvelopeArrayDesign} we offer some qualitative intuition for using interferometry to measure monopole signals and provide rules-of-thumb for the design of a global signal interferometer. Section \ref{sec:MathForm} establishes a general framework for data analysis, providing a convenient language for considering various data analysis trade-offs and choices. In Section \ref{sec:SimResults} we perform numerical simulations to compare the performance of single-element experiments and interferometers, before summarizing our conclusions in Section \ref{sec:Conc}.


%

\section{Back-of-the-envelope array design}
\label{sec:BackOfEnvelopeArrayDesign}

In this section, we use simplified toy models to build intuition for the types of interferometer arrays that are best suited for probing the global signal.  The goal here is to provide a rough sense for what might be a sensible design, which we will analyze in a more numerically detailed fashion in the rest of the paper.

\subsection{A semi-qualitative picture of a global signal interferometer}
\label{sec:PictorialToyIntro}
\begin{figure}[t]
	\centering
	\includegraphics[width=0.50\textwidth] {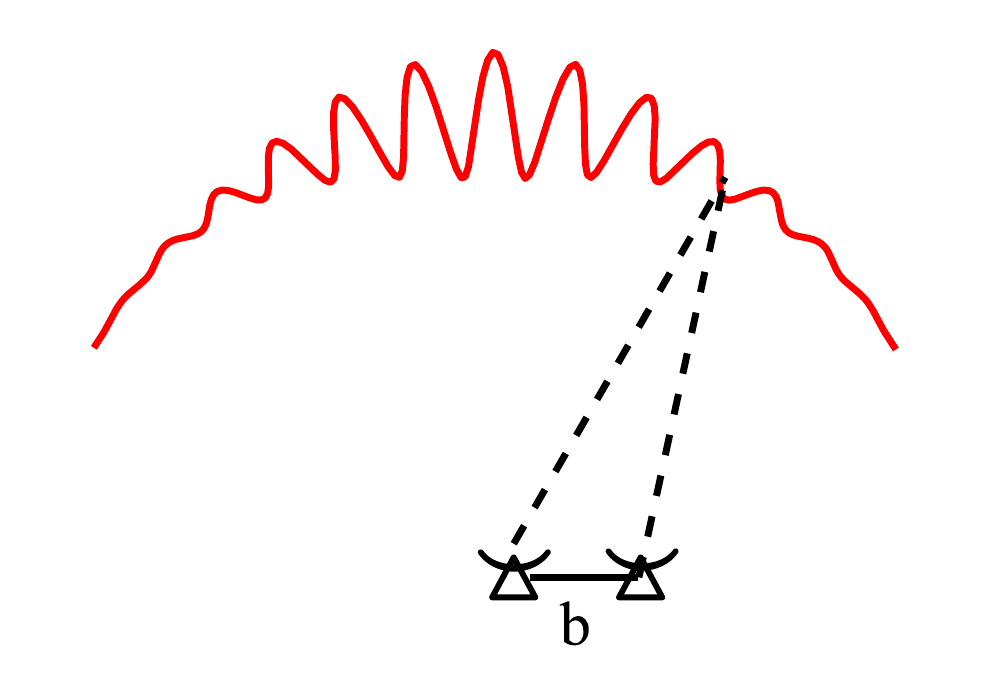}
	\caption{A schematic illustration of a two-element interferometer. The two antennas, separated by a baseline length $\b$, measure a fringe pattern on the sky, shown as a cosine function on an arc. This pattern would integrate to zero over the sky if there were no primary beam and the sky were infinite in extent. However, both these assumptions are violated in practice: the curved sky is finite, and the antennas also produce a beam pattern that attenuates sensitivity to certain parts of the sky (e.g., the horizon for zenith-pointing elements). The result is a pattern that does not integrate to zero and an interferometer that is sensitive to the monopole mode.}
	\label{fig:beam_pattern_cartoon}
\end{figure}

Consider first a purely qualitative picture of a two-element interferometer. An interferometer returns a visibility by integrating over a primary-beam attenuated fringe pattern on the sky. If the primary beam did not exist (i.e., if it were constant over the entire sky) and the sky were infinite in extent, it would be impossible to measure a monopole. With those approximations, the fringe pattern would be purely sinusoidal, and integrating such a pattern against a constant (monopole) mode would return zero. However, as shown schematically in Figure \ref{fig:beam_pattern_cartoon}, the enveloping presence of the primary beam and the finite extent of the sky prevent the interferometer's response from integrating to zero. This allows an interferometer to be sensitive to the monopole mode.

\begin{figure}[h]
	\centering
	\includegraphics[width=0.50\textwidth] {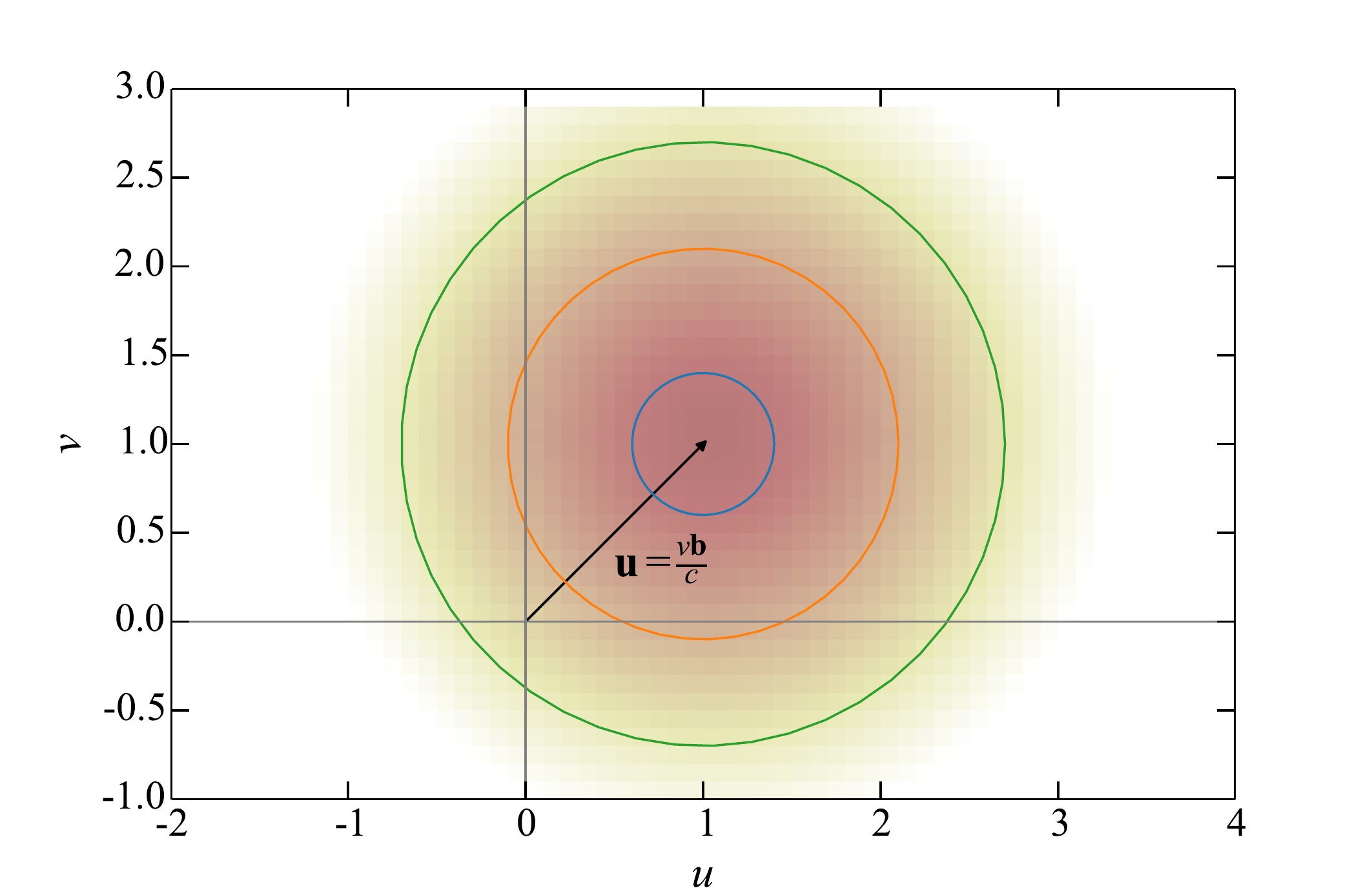}
	\caption{A schematic illustration of the $uv$ plane for an interferometer. The $uv$ plane is the Fourier dual to image space, and an interferometer with baseline $\b$ makes a measurement at the point $\mathbf{u} = \nu\b / c$. Measuring the monopole signal requires making a measurement at the origin of the $uv$ plane, which is probed by a baseline of zero length. This seems impossible without measuring the auto-correlation of an antenna with itself, until one recalls that a real interferometer does not measure an exact point: the antenna's beam has a $uv$ plane footprint that smears the measurement, so that the measurement contains information from nearby $uv$ points. Since a narrower beam corresponds to a larger convolving kernel, two narrow-beam antennas placed close together would make a measurement that incorporates information from the monopole, as shown by the green circle in the figure. Antenna size scales both the center point and the extent of the uv kernel, thus leaving the response to the global signal approximately the same. The scales on the $uv$ axes shown here are thus somewhat arbitrary. \acl{Added last two sentences}}
	\label{fig:uv_cartoon}
\end{figure}

An equivalent way to look at the problem is to move into Fourier space and to examine the $uv$ plane, as shown in Figure \ref{fig:uv_cartoon}. An interferometer with baseline length $\b$ will measure the sky at the point $\mathbf{u} = \nu\b/c$ in the $uv$ plane. To measure the monopole, one must recover information from the origin, which corresponds to a zero-length baseline. This is possible because multiplying the sky by the primary beam in image space corresponds to convolving by the primary beam's $uv$ plane footprint in Fourier space. This footprint is typically on the order of the physical size of the antenna element, measured in wavelengths. Its effect is to smear out the point of measurement on the $uv$ plane (i.e. the measurement incorporates information from other nearby $uv$ locations), with narrower image-space beams corresponding to larger smears. As shown by the green circle in Figure \ref{fig:uv_cartoon}, if the baseline is short enough and the image-space beam is narrow enough, then the measured signal will include information from the global signal.

At first sight, the result that an interferometer can be sensitive to the monopole of the sky seems to contradict standard ideas in interferometry. To help clarify this point of confusion,
let us first review the standard lore. The van Cittert-Zernicke Theorem states that the cross-correlation
between the electric field at locations $\mathbf{x}_1$ and $\mathbf{x}_2$ is given by
\begin{equation}
\langle E(\mathbf{x}_1) E(\mathbf{x}_2)^* \rangle \propto \int d\Omega \, T(\rhat) \exp \left[ - i 2\pi \frac{\nu}{c} \rhat \cdot (\mathbf{x}_1 - \mathbf{x}_2 ) \right],
\end{equation}
where we have opted to describe the sky in terms of its brightness temperature rather than the specific intensity, in order to conform to our convention in later sections. With an interferometric baseline, one correlates
not the electric field at two specific points, but rather, the integrated electric fields over the physical
areas of the antennas forming the baseline. We therefore measure
\begin{equation}
V \propto \int d^2 \mathbf{x}_1 d^2 \mathbf{x}_2 f (\mathbf{x}_1) g(\mathbf{x}_2) \langle E(\mathbf{x}_1) E(\mathbf{x}_2)^* \rangle,
\end{equation}
where we have assumed that the antennas are coplanar, allowing us to integrate only over two-dimensional
versions of $\mathbf{x}_1$ and $\mathbf{x}_2$. The functions $f$ and $g$ describe the electric field sensitivities of the first and second antennas comprising the baseline, respectively. They are assumed to be equal to zero outside the rough physical extent of the
antennas. Suppose the sky emission is constant (i.e., consisting only of a monopole), so that $T(\rhat) = T_0$. Combining our expressions with this restriction and the explicit notation $\rhat = (l,m,n)$, we obtain
\begin{equation}
V \propto T_0 \int \frac{dl dm}{\sqrt{1- l^2 - m^2}} \,\tilde{f}\!\left( l \nu / c , m \nu / c \right) \tilde{g}^* ( l \nu / c , m \nu / c ),
\end{equation}
where
\begin{equation}
\tilde{f}\left( \frac{l \nu}{c}, \frac{m \nu}{c}  \right) \equiv \int dx dy f(x,y)e^{ - i 2 \pi \frac{\nu}{c} (lx + my)}
\end{equation}
is the two-dimensional Fourier transform of $f$, and similarly for $\tilde{g}$, with $\mathbf{x} = (x,y)$.

Now, suppose we operate under the flat-sky approximation, which as we shall see, is inappropriate for global signal interferometry. This is equivalent to omitting the
factor of $\sqrt{1-l^2 - m^2}$ in the denominator of our expression for the visibility $V$.
If one further extends the limits of the integration to infinity, Parseval's theorem applies and
one obtains
\begin{equation}
\label{eq:fdotg}
V\Bigg{|}_\textrm{flat, finite} \propto T_0 \int d\mathbf{x} f (\mathbf{x}) g(\mathbf{x}).
\end{equation}
If the two antennas that form our baseline do not overlap, then neither will $f$ and $g$, resulting
in $f (\mathbf{x}) g(\mathbf{x}) = 0$ and thus $V=0$. This is the standard result that suggests
that it is impossible to measure the monopole with an interferometer. Phrased differently, the
physical size of the antennas ($f$ and $g$) make it difficult to have a baseline short enough
for there to be substantial overlap with the $\mathbf{u} = 0$ mode in Figure \ref{fig:uv_cartoon}.

With the full expression for $V$, however, one sees that the response to the monopole does
not necessarily vanish, even if the antennas are not physically co-located. Abandoning the
assumptions of a flat, finite sky, one may define
\begin{equation}
\label{eq:FspaceEffectiveAperture}
\widetilde{F} (l, m, \nu) \equiv
\begin{cases}
\frac{\tilde{f} (l\nu / c,m\nu / c)}{(1-l^2 - m^2)^{1/4}} \quad & \textrm{if} \quad l^2 + m^2 < 1 \\
0 \quad &\textrm{otherwise},
\end{cases}
\end{equation}
and similarly for $g$. Parseval's theorem can then be applied to our expression for $V$, despite
the complications of a curved, finite sky. The result (suppressing the frequency dependence for
notational simplicity) is
\begin{equation}
\label{eq:effFdotG}
V \propto T_0 \int d\mathbf{x} F(\mathbf{x}) G(\mathbf{x}),
\end{equation}
and does not vanish because the effective apertures $F$ and $G$ will in general overlap. \emph{In words,
an interferometer is sensitive to the monopole mode because a monopole does not appear as a
constant as far as the interferometer is concerned. From the interferometer's viewpoint, the finite
extent of the sky means that it is effectively making a measurement on an infinite plane, but one that is identically zero beyond a circle of fixed radius given by the horizon.} The interferometer therefore
sees a constant sky as having spatial structure, allowing for a non-zero response. This response
is further enhanced by the geometric effect of projecting a spherical hemisphere down to a two-dimensional
image plane. Such a projection maps large solid angles near the horizon to small portions of the image
plane, leading to an ``edge brightening" effect, yet again imprinting spatial structure on the signal so that it can be picked up by an interferometer. These
effects and their importance for an interferometric measurement of the global signal were independently
noted in \citet{Thyagarajan_et_al2015}. 

\begin{figure}[t]
	\centering
	\includegraphics[width=0.50\textwidth] {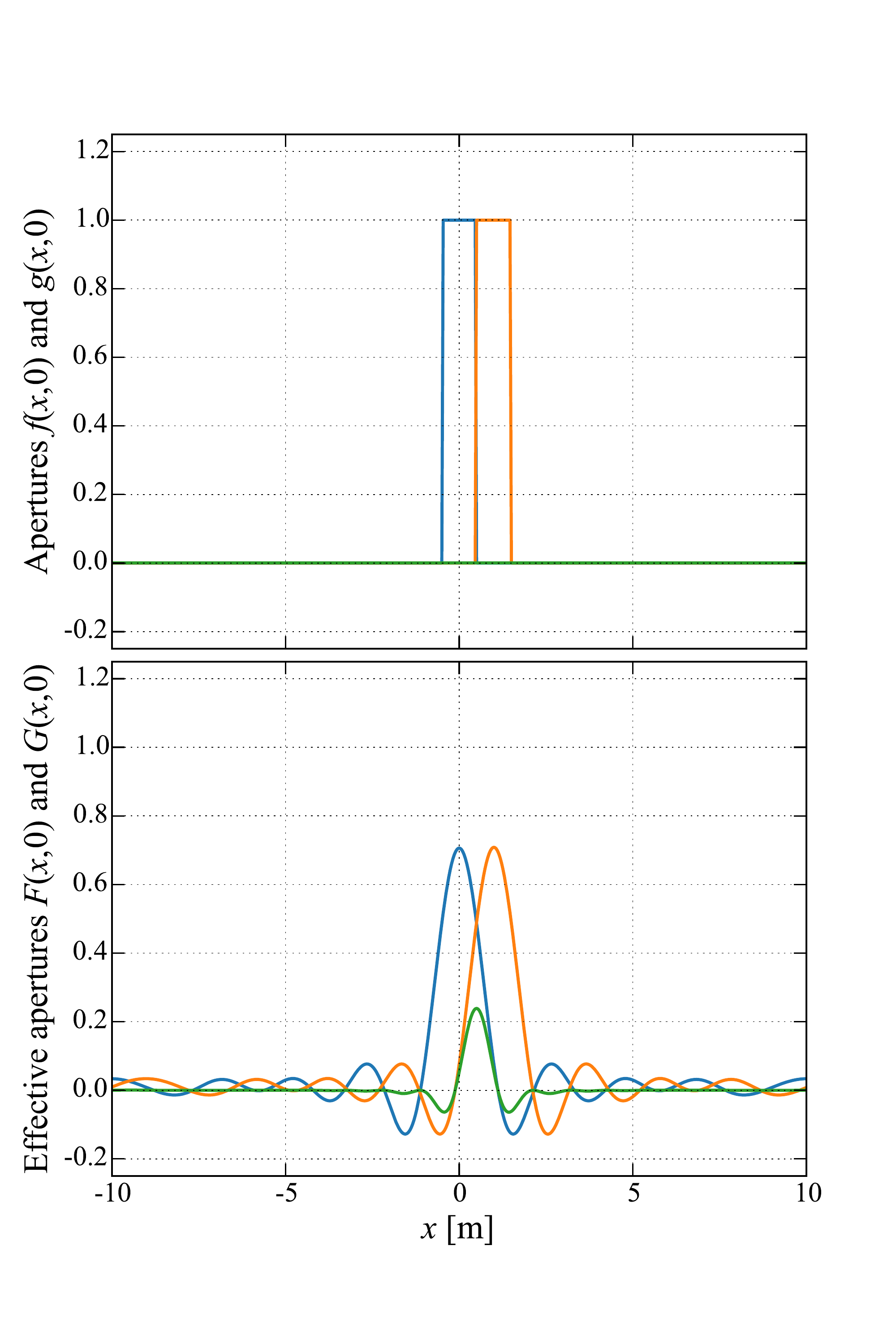}
	\caption{Top: Cross-section of two circular apertures (shown in blue and orange) of radius $0.5\,\textrm{m}$ placed
	side-by-side, with their product (identically zero everywhere) in green. Bottom: Cross-section of the corresponding effective apertures at $150\,\textrm{MHz}$, with the (now non-zero) product again shown in green.
	Whereas the lack of overlap between the two apertures in the top panel implies that
	interferometers are insensitive to the monopole, this conclusion is incorrect when finite
	curved sky effects are taken into account. Incorporating such effects, the effective
	apertures overlap and give rise to a non-zero response to the monopole, which can be obtained by integrating the green curve.}
	\label{fig:effAperture}
\end{figure}

\begin{figure}[h]
	\centering
	\includegraphics[width=0.50\textwidth] {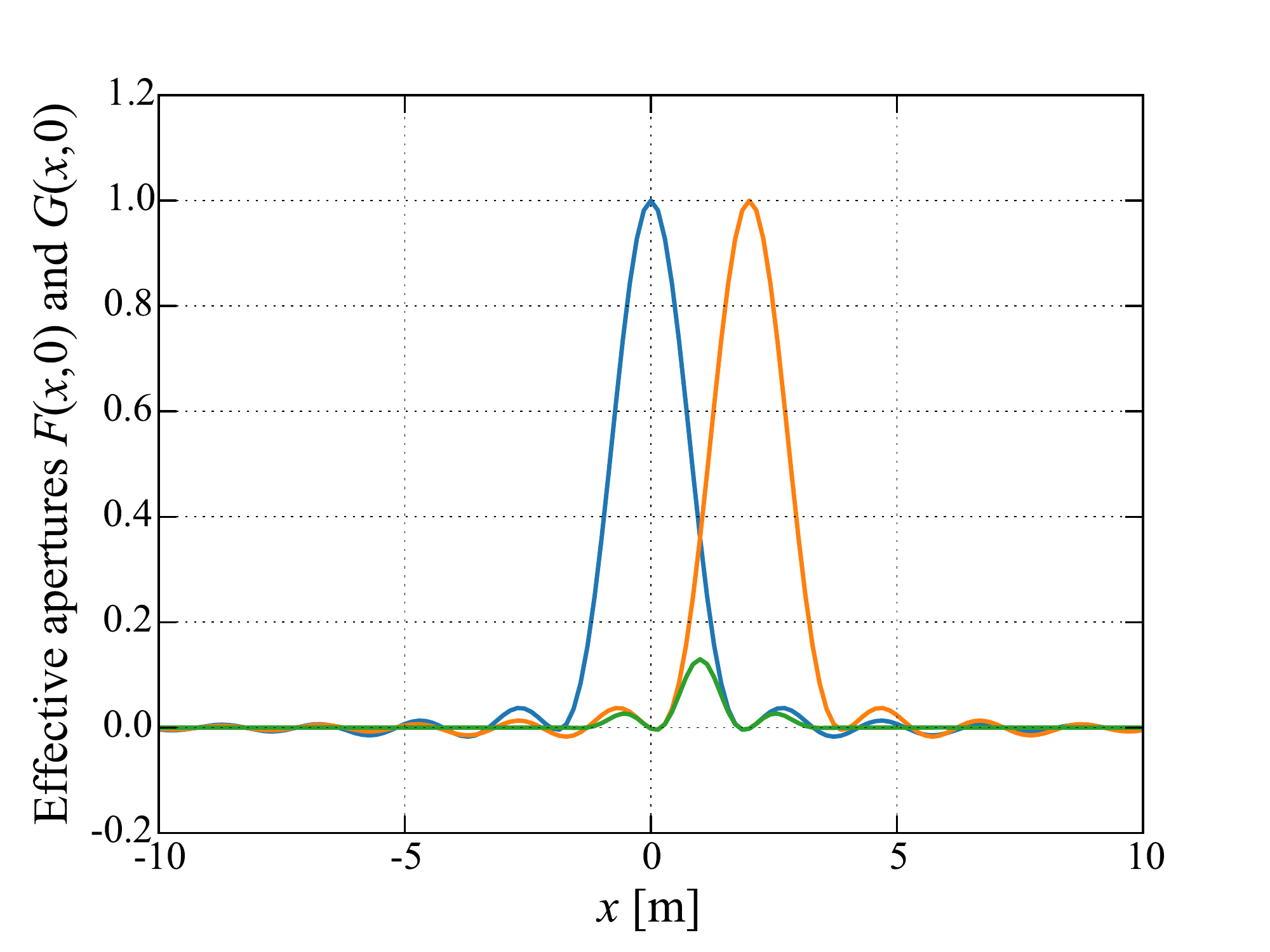}
	\caption{Similar to the bottom plot in Figure \ref{fig:effAperture}, but using realistic beam models from the PAPER array. The antennas are once again placed side-by-side. The results shown here are the effective apertures for the two antennas at $150\,\textrm{MHz}$, arbitrarily peak-normalized to unity. Again, the overlap implies a sensitivity to the monopole, with the response being proportional to the integral over the product of the two effective apertures (shown in green).}
	\label{fig:effAperture_PAPER}
\end{figure}

We demonstrate our results numerically with a toy example in Figure \ref{fig:effAperture}. The top panel
shows the cross-sections of two circular apertures that each have radius $0.5\,\textrm{m}$. The apertures are packed as
closely together as possible without overlapping, and in the infinite, flat-sky approximation, Equation
\eqref{eq:fdotg} implies that an interferometric baseline formed by pairing these two apertures will
have no sensitivity to the monopole mode. In the bottom panel are the effective aperture functions
(i.e., the correct ones to use when considering the curved, finite sky) at $150\,\textrm{MHz}$ for the same configuration. The effective apertures clearly overlap, and consequently Equation \eqref{eq:effFdotG} implies a sensitivity to the monopole.\footnote{The bottom panel of Figure \ref{fig:effAperture} looks superficially like the Fourier transform of the top panel, but we emphasize that this is \emph{not} what is being plotted. To obtain the bottom panel, one Fourier transforms the top panel, applies the truncation and edge brightening implied by Equation \eqref{eq:FspaceEffectiveAperture}, and then inverse Fourier transforms back to the original space, yielding the effective aperture functions used in Equation \eqref{eq:effFdotG}.} 
Another example is provided in Figure \ref{fig:effAperture_PAPER}, where we move beyond a toy model and use realistic antenna models \citep{JonnieBeam} from the Donald C. Backer Precision Array for Probing the Epoch of Reionization (PAPER; \citealt{Parsons2010}). Physically, the PAPER antennas are $\sim 2 \,\textrm{m}$ in extent. Placing two antennas as close as possible results in the effective aperture functions shown in Figure \ref{fig:effAperture_PAPER}. Again, the non-zero overlap results in sensitivity to the monopole mode.

In the rest of Section \ref{sec:BackOfEnvelopeArrayDesign} we will make this rough picture more mathematically precise and develop an array design that not only maximizes sensitivity to the global signal, but also ensures that subsequent foreground subtraction operations are robust. This will require being cognizant of the data analysis strategy. This strategy, which we will develop fully in Section \ref{sec:MathForm}, first analyzes the visibilities on a frequency-by-frequency basis to estimate the strength of the monopole mode at each frequency. The result is a spectrum whose smooth components are then removed in an attempt to remove foregrounds. Essential to the success of this removal is the assumption that the instrument does not impose complicated frequency structures on the true sky spectra. This provides a strong constraint on our instrumental design. In service of this goal, when analyzing interferometric data we will discard data from the auto-correlation of an antenna itself, in order to avoid the noise bias effects that pose extra challenges in calibration of single-element experiments.

\subsection{Sparse or packed arrays?}

Suppose we consider an array consisting of just a single baseline and ask what baseline length $b$ optimizes recovery of the global signal.  Intuitively, a short $b$ increases sensitivity to the signal, since (for a given $uv$ plane primary beam kernel) short baselines have the greatest overlap with the $\mathbf{u}=0$ mode.  On the other hand, foregrounds contamination is worst for short baselines, since they are mostly sensitive to the smoothest spatial modes of the sky, where foregrounds dominate.  There must therefore exist an intermediate baseline length that best balances these two competing demands, which we will now compute.

In the flat-sky approximation, the visibility response $V(\mathbf{b})$ of a baseline $\mathbf{b}$ to the sky temperature $T(\boldsymbol \theta)$ is given by
\begin{equation}
\label{eq:Vb}
V(\mathbf{b}) = \int  T(\boldsymbol \theta) A(\boldsymbol \theta) \exp \left( -i 2 \pi \frac{\nu}{c} \mathbf{b} \cdot \boldsymbol \theta \right) d^2 \mathbf{\theta},
\end{equation}
where $ A(\boldsymbol \theta)$ is the primary beam pattern. Without loss of generality, we may normalize our primary beam such that $A(\mathbf{0}) = 1$. In principle, our use
of the flat-sky approximation is inappropriate for a discussion of global signal interferometry,
given the conceptual picture we presented in the previous section. However, we will
invoke the flat-sky approximation only for the purposes of enhancing physical intuition, and
the formalism and numerical results of subsequent sections will properly incorporate the
curved sky. In fact, if one prefers, one may simply replace all the aperture functions
(which enter through the beam patterns) with the effective aperture functions. The
rest of the mathematics---including what we do in this section---then carries through without change.\acl{Added to the end of this paragraph}

For notational compactness, we will not explicitly highlight the frequency dependence of $T_0$, $A(\boldsymbol{\theta})$, and $V(\mathbf{b})$, although of course these quantities all implicitly vary with frequency. Setting $T(\boldsymbol \theta) = T_0$ for a monopole signal, one obtains the result
\begin{equation}
\label{eq:blMonoResponse}
V(\mathbf{b}) = T_0 \widetilde{A} \left( \nu \mathbf{b} / c \right),
\end{equation}
with $\widetilde{A}$ is defined as the Fourier transform\footnote{Throughout this paper, we adopt a Fourier convention where $f(\boldsymbol \theta)$ has a Fourier transform $\widetilde{f}(\mathbf{u}) \equiv \int f(\boldsymbol \theta) e^{-i 2 \pi \mathbf{u} \cdot \boldsymbol \theta  } d^2 \boldsymbol \theta.$ The inverse transform is correspondingly given by $f(\boldsymbol \theta) = \int \widetilde{f}(\mathbf{u}) e^{i 2 \pi \mathbf{u} \cdot \boldsymbol \theta  }d^2 \mathbf{u}.$ \mep{make sure this stays with rest}} of $A(\boldsymbol \theta)$. This suggests that an appropriate (though not necessarily optimal) estimator $\widehat{T}_0$ for the global signal $T_0$ might be
\begin{equation}
\label{eq:singleBlSillyEst}
\widehat{T}_0 = \frac{V( \mathbf{b})}{\widetilde{A} \left( \nu \mathbf{b} / c \right)}.
\end{equation}
Intuitively, a baseline $\mathbf{b}$ has sensitivity $\widetilde{A} \left( \nu \mathbf{b} / c - \mathbf{u} \right)$ to spatial wavenumber $\mathbf{u}$, so the prescription suggested here is to simply divide the measured visibility by the response to the $\mathbf{u}=0$ mode. In the absence of foregrounds and noise, this is guaranteed to return the true $T_0$.  In reality, of course, we have contributions from both foregrounds and noise.  To describe the former, we can write $T(\boldsymbol \theta)$ as the sum of $T_0$ and a foreground contribution $T_\textrm{fg} (\boldsymbol \theta)$.  This then yields a foreground perturbation $V_\textrm{fg} (\mathbf{b})$ to the visibility, of the form
\begin{equation}
V_\textrm{fg} (\mathbf{b}) = \int  \widetilde{T}_\textrm{fg} (\mathbf{u}) \widetilde{A} \left( \frac{\nu \mathbf{b}}{c} - \mathbf{u} \right) d^2 u,
\end{equation}
where we have applied the convolution theorem to Equation \eqref{eq:Vb}, with $\widetilde{T}_\textrm{fg}$ denoting the Fourier transform of the foreground temperature field.  With this perturbation to $V(\mathbf{b})$, our estimator contains more than just the contribution from the true global signal:
\begin{equation}
\label{eq:PerturbedEst}
\widehat{T}_0 = T_0 + \frac{1}{\widetilde{A} \left( \nu \mathbf{b} / c \right)} \left[ \int  \widetilde{T}_\textrm{fg} (\mathbf{u}) \widetilde{A} \left( \frac{\nu \mathbf{b}}{c} - \mathbf{u} \right) d^2 u + n \right],
\end{equation}
where we have included an additive instrumental noise contribution $n$ to the visibility.  Taking the ensemble average of both sides and assuming that the noise averages to zero (i.e., there are no persistent instrumental systematics such as crosstalk), it follows that the average deviation $\Delta T_0$ from the truth is given by
\begin{equation}
\label{eq:Deviation}
\Delta T_0 \equiv \langle \widehat{T}_0 \rangle - T_0 \approx \frac{\widetilde{T}_\textrm{fg} (\nu \mathbf{b} / c)}{\widetilde{A} \left( \nu \mathbf{b} / c \right)} \int   \widetilde{A} \left( \frac{\nu \mathbf{b}}{c} - \mathbf{u} \right) d^2 u,
\end{equation}
where we have assumed that the primary beam $A$ is a relatively broad function on the sky, resulting in a compact $uv$ plane footprint $\widetilde{A}$.  This allows the factor of $\widetilde{T}_\textrm{fg}$ in Equation \eqref{eq:PerturbedEst} to be evaluated at $\mathbf{u} = \nu \mathbf{b} / c$ and factored out of the integral. What remains in the integral in Equation \eqref{eq:Deviation} is simply the integral of the primary beam kernel over the entire $uv$ plane, which equals $A(\mathbf{0}) = 1$. We thus have
\begin{equation}
\Delta T_0 = \frac{\widetilde{T}_\textrm{fg} (\nu \mathbf{b} / c)}{\widetilde{A} \left( \nu \mathbf{b} / c \right)}.
\end{equation}
This represents the bias that foregrounds introduce into our estimate of the global signal, which we can seek to minimize by varying $\mathbf{b}$.

For illustrative purposes, let us consider specific models for $\widetilde{A}$ and $\widetilde{T}_\textrm{fg}$.  If the primary beam is taken to be a 2-D Gaussian with a width of $\theta_b^2$, then our normalization convention dictates that $\widetilde{A}(\mathbf{u}) = 2 \pi \theta_b^2 \exp (-2 \pi^2 \theta_b^2 u^2 )$, where $u \equiv |\mathbf{u}|$.  As for $\widetilde{T}_\textrm{fg}$, one can imagine the foregrounds to have statistical properties described by some angular power spectrum $C_\ell$.  Often, $C_\ell$ is fit by a power law so that $C_\ell \propto \ell^{-\alpha}$, where $\alpha$ is typically between $2$ and $3$ (depending on various factors such as frequency and Galactic latitude).  In the flat-sky approximation, we have $\ell \sim 2 \pi u$, which allows us to translate the angular power spectrum into a power spectrum $P(\mathbf{u}) \propto u^{-\alpha}$ on the $uv$ plane.  Given this, it is reasonable (on dimensional grounds) to take $\widetilde{T}_\textrm{fg} (u)\propto u^{-\alpha/2}$, which yields
\begin{equation}
\label{eq:DeviationPowGauss}
\Delta T_0(b) \propto \left( \frac{\nu b}{c}\right)^{-\alpha/2} \exp \left(2 \pi^2 \theta_b^2 \frac{\nu^2 b^2}{c^2}  \right),
\end{equation}
where $b \equiv | \mathbf{b} |$.  Minimizing this expression by differentiating with respect to $b$ gives an optimal baseline length $b_\textrm{opt}$ of
\begin{equation}
\label{eq:b_opt}
\frac{b_\textrm{opt}}{\lambda} = \frac{1}{2\pi \theta_b} \sqrt{\frac{\alpha}{2}}.
\end{equation}
Based on the discussion of aperture sizes in Section \ref{sec:PictorialToyIntro}, we immediately
recognize the factor of $(2 \sqrt{2} \pi \theta_b)^{-1}$ as being the characteristic ``radius" of a receiving element in units of wavelength. Now, the remaining factor $\sqrt{\alpha}$ is of order unity and less than
$2$. Since one cannot place antenna elements closer than their diameters,\footnote{Note that
when dealing with elements like dipole antennas (as opposed to aperture-like elements such as dishes), the electrical response can be larger than the physical antenna size. With careful antenna design, it may therefore be possible to abide by Equation \eqref{eq:b_opt}. \acl{Just a marker to remind myself that I added this, and changed the text surrounding the footnote.}} we cannot in fact
achieve the optimal baseline length suggested by Equation \eqref{eq:b_opt}. However, we can
come close to this by placing the antennas as close together as is physically possible. Essentially, our
optimization suggests such a compact configuration because the foreground power does not decrease
dramatically with increasing spatial wavenumber (i.e., we never have $\alpha \gg 1$), so the reduced sensitivity to the global signal from having a longer baseline is not worth the relatively small decrease in foreground contamination.

In the derivation that we have just presented, we focused exclusively on minimizing the systematic \emph{bias} that would result from foreground contamination.  Alternatively, we could have instead chosen to minimize the \emph{variance} (i.e., the error bars) on our estimator $\widehat{T}_0$.  Unlike the bias, the variance contains a noise term, since $\langle n \rangle = 0$ in the absence of systematics, but $\langle |n|^2 \rangle$ will be non-zero.  This will tend to reduce the optimal baseline length, given that short baselines increase signal-to-noise.  But since Equation \eqref{eq:b_opt} predicts close to the shortest possible baseline anyway, our minimum-bias solution also serves as an excellent approximation to a minimum-variance solution.

Making a slight leap from a single baseline to a full interferometer array, this section argues for a packed array, where antenna elements are placed as close together as possible. A packed array naturally results in a regular, periodic arrangement of antennas, giving a large number of identical copies of our single (short) baseline. Our conclusion then rests on the assumption that a large regular array is essentially just that---a large collection of repeated, short baselines---and no more. In general, this is not a good description of an array, since for large arrays, even close-packed antenna configurations provide longer baselines that might provide valuable information about foregrounds for advanced data analysis techniques. In terms of the cosmological global signal, however, longer baselines have very little sensitivity to the signal of interest. We can see this from Equation \eqref{eq:blMonoResponse}, where $\widetilde{A}$ is typically a function that drops off away from the origin, so that as one increases $\mathbf{b}$ from zero (i.e., a single element experiment) to a short baseline to a long baseline, the visibility response to the monopole $T_0$ drops. For measuring the signal, long baselines therefore contribute negligibly, and a large array can be thought of as simply a large collection of multiple short baselines. We can therefore make the leap from the single baseline derivations of this section to argue that packed arrays are desirable.
%

\subsection{Wide beams or narrow beams?}
\label{sec:beamSize}

The arguments in the previous subsection suggested a particular \emph{relative} placement of antenna elements: antennas should be packed together as tightly as possible. However, the \emph{absolute} scale of the array remains unspecified. Primary beams smaller than $\sim1-2^\circ$ will likely have difficulty distinguishing a representative $21\,\textrm{cm}$ global signal from local fluctuations \citep{BittnerLoeb2011}. However, that constraint still leaves a large range of potential primary beam sizes and baseline lengths. \mep{Updated for reviewer comment} In this section, we answer the question of whether it is better to have a packed array with physically small antenna elements (and therefore short baselines and wide primary beams), or a packed array with larger elements (and therefore longer baselines and narrow primary beams). We will ultimately find that although narrowest beams on longest baselines maximize raw foreground reduction, they also introduce spectral ripples that are difficult to remove. Hence, we will find intermediate beam and baseline sizes to be optimal.

As a first guess, one could imagine inserting our expression for $b_\textrm{opt}$, Equation \eqref{eq:b_opt}, into Equation \eqref{eq:DeviationPowGauss} to yield an equation whose only free parameter is the primary beam size $\theta_0$. Minimizing this equation by varying the beam size then suggests that the beam ought to be made as small as possible. However, since any discussion of an array's absolute size will necessarily tie the array to absolute angular scales on the sky, a more nuanced discussion of foreground properties is required beyond the set-up in the previous subsection, which only required that the angular power spectrum of foregrounds was monotonically decreasing.

One important property of the foreground sky is the fact that it is not rotationally invariant---the galactic plane, for example, is far brighter than the galactic poles. This is not captured by the angular power spectrum of foregrounds, which abuses the notion of a power spectrum by assuming statistical isotropy for a sky that is clearly anisotropic. A global signal experiment with a narrow beam can take advantage of cooler regions in the galaxy, selectively observing only where foregrounds are known to be dimmer, since the cosmological global signal is by definition the same everywhere on the sky. The narrower the primary beam, the more selectively one can implement such an angular foreground avoidance scheme, and the lower the foreground contamination. Arrays with narrow primary beams, large antennas, and long baselines therefore see dimmer foregrounds for two reasons: the narrow primary beam allows for cleaner selections of cool patches of the sky, and the necessarily longer baselines also sample foregrounds on finer angular scales (higher $\ell$), which are weaker because $C_\ell$ is a decreasing function for galactic foregrounds.

Narrow primary beams, however, are not without their drawbacks. Angular avoidance strategies alone cannot mitigate foregrounds to the level required for a detection of the cosmological global signal. An angular avoidance strategy in principle allows the rejection of any foregrounds that are not spatially constant (i.e., are not the monopole), but are unable to remove the monopole component of foregrounds. Put another way, an observational strategy that avoids the strongest foregrounds will reduce the magnitude of foreground contamination considerably, but will at best only be able to reduce the contamination to the minimum foreground temperature on the sky, which can still be much brighter than the cosmological signal. Ultimately, one must therefore also rely on spectral foreground subtraction methods, and this is where narrow beams and long baselines may not be advantageous. Spectral subtraction typically exploits the intrinsic smoothness of foreground spectra, projecting out smooth components of the data. For such a procedure to be successful, one must avoid having an instrument that imprints extra spectral features into the data. Long baselines are particularly prone to such imprints, since the angular mode number $\ell \sim 2 \pi u \sim 2 \pi b / \lambda$ probed by a baseline $b$ varies more rapidly with frequency (or wavelength) when $b$ is large, allowing non-uniform spatial features of the sky to couple more strongly into spectral ripples.\footnote{Indeed, this is a common concern for $21\,\textrm{cm}$ tomography, and is the origin of the ``foreground wedge" signature seen in power spectrum measurements \citep{DattaWedge,MoralesWedge,CathWedge,VedanthamWedge,AaronDelay,NithyaWedge,JonnieWedge,meWedge1,meWedge2}. However, one key difference is that most interferometer arrays targeting the power spectrum are not tightly packed (though they do tend to be quite compact). Such arrays are therefore not subject to our constraint that the primary beam width and the baseline length vary in a strictly reciprocal fashion. One exception to this is the Hydrogen Epoch of Reionzation Array (HERA), which does have close-packed elements \citep{JonnieHERA}.} Such spectral ripples will survive a spectral foreground subtraction that projects out smooth modes (although see \citealt{Liu_Switzer_2014} for a proposal for how these ripples can be modeled \emph{in situ} from the data itself), leaving residuals that may be indistinguishable from the cosmological signal. Combining this with our previous discussion, we see that an array with longer baselines and narrower primary beams may see dimmer foregrounds prior to spectral foreground subtraction, but may imprint spectral signatures that result in greater post-subtraction foreground residuals. An optimal array is one with a beam size that is chosen to balance these two competing demands.

Because spatial features of the foreground sky such as the Galactic plane are difficult to model statistically, numerical simulations are required to find the right balance in primary beam size. To perform such simulations, we first form simulated foreground skies between $100$ and $200\,\textrm{MHz}$ by extrapolating the $408\,\textrm{MHz}$ Haslam map \citep{Haslam_408MHz_map} pixel-by-pixel using a power-law-like relation
\begin{equation}
\label{eq:HaslamExtrap}
T(\nu) = T_\textrm{Haslam} (\rhat) \left( \frac{\nu}{408\,\textrm{MHz}}\right)^{\alpha_0 + \sum_{n=1}^{3} \alpha_n \left[ \ln \left(\nu / \nu_0 \right) \right]^n},
\end{equation}
where $T_\textrm{Haslam} (\rhat)$ represents the Haslam map and $\alpha_0$ is held fixed at $-2.5$ \citep{Liu_21cm_Fg}, whereas $\alpha_1$, $\alpha_2$, and $\alpha_3$ are drawn pixel-by-pixel from zero-mean Gaussian distributions with standard deviations of $0.1$, $0.03$, and $0.01$, respectively. Conceptually, higher curvature components to the spectrum are given less weight, in accordance with the empirical eigenmode analysis of \citet{angelicaGSM} and its mathematical interpretation in \citet{Liu_21cm_Fg}. It is important that our numerical simulations are based on extrapolations with pixel-to-pixel variations. This will make the numerical explorations of Section \ref{sec:Hchoices} more realistic. There, we incorporate the Haslam map as a model of the foreground sky as part of our data analysis pipeline (in an attempt to suppress foregrounds), which in reality will differ somewhat from the true sky. Pixel-to-pixel variations give rise to low-frequency maps that roughly look like the Haslam map, but with slight differences in their details, reflecting the imperfections of any sky template we may choose to use.

We assume that observations are centered on the Northern Galactic Pole (NGP) with the extent of the field defined by the primary beam of the instrument. We take the primary beam to be a Gaussian attenuated by a cosine (to ensure that the primary beam vanishes at the horizon):
\begin{equation}
\label{eq:TaperedGauss}
A(\theta, \varphi) = \exp \left( -\frac{1}{2} \frac{\theta^2}{\theta_b^2} \right) \cos \theta,
\end{equation}
where $\theta_b$ controls the width of the primary beam. We assume (rather conservatively) that the beam width is proportional to the observation wavelength, and subsequent quotations of $\theta_b$ in this section refer to the beam width at the \emph{lowest} frequency of observation. To measure the global signal, we compute
\begin{equation}
\label{eq:NormedSimpleEst}
\widehat{T}_0 (\nu) = \frac{ \sum_j \left[ \int  A(\rhat, \nu) \exp\left(i 2\pi \frac{\nu }{c}\mathbf{b}_j \cdot \rhat \right)d\Omega \right] V(\mathbf{b}_j, \nu)}{ \sum_k \big{|} \int  A(\rhat, \nu) \exp\left(i 2\pi \frac{\nu }{c}\mathbf{b}_k \cdot \rhat \right)d\Omega \big{|}^2}.
\end{equation}
Although we defer a full discussion of data analysis to Section \ref{sec:MathForm}, we can understand the essential features of this estimator as a generalization of Equation \eqref{eq:singleBlSillyEst}. First, this estimator does not require the flat-sky approximation. In addition, it incorporates a signal-to-noise weighting of measurements from different baselines. To see this, note that the term in the numerator enclosed by the square brackets is precisely the visibility response of a baseline to a monopole sky of unit amplitude. Baselines with a greater response are given greater weight as visibilities from different baselines are summed together, before normalizing the final result. If the array consists of a single baseline, the summations in both the numerator and denominator disappear, and the estimator reduces to Equation \eqref{eq:singleBlSillyEst} once the flat-sky approximation is invoked.

Following an initial estimate of the sky spectrum, we further suppress foregrounds by fitting a polynomial to the logarithm of the spectrum. Subtracting off the smooth polynomial fit, one obtains a residual spectrum
\begin{equation}
\label{eq:FgFit}
T_\textrm{res} (\nu) = \widehat{T}_0(\nu) -  \exp \left[ \sum_{n=0}^{N_\textrm{poly}} a_n p_n( \log \nu) \right],
\end{equation}
where $p_n$ denotes the $n$th Legendre polynomial, with a corresponding expansion coefficient $a_n$ obtained from fitting $\log \widehat{T}_0$ up to order $N_\textrm{poly}$. The set of polynomials that one fits to is arbitrary, and our choice of Legendre polynomials is simply one of convenience. In fact, there is nothing sacred about polynomial subtraction, and alternatives such as principal component analyses \citep{Liu_21cm_Fg,L13,Liu_Switzer_2014} are worth exploring. Whatever foreground model is ultimately used, one must simply take care to ensure that any possible loss of cosmological signal (resulting from the subtraction of spectral modes that the foregrounds and the signal have in common) is accurately quantified. We will do so in Section \ref{sec:SimResults}, when we propagate numerical simulations of fiducial instruments through to astrophysical parameters.

\begin{figure}[h]
	\centering
	\includegraphics[width=0.50\textwidth] {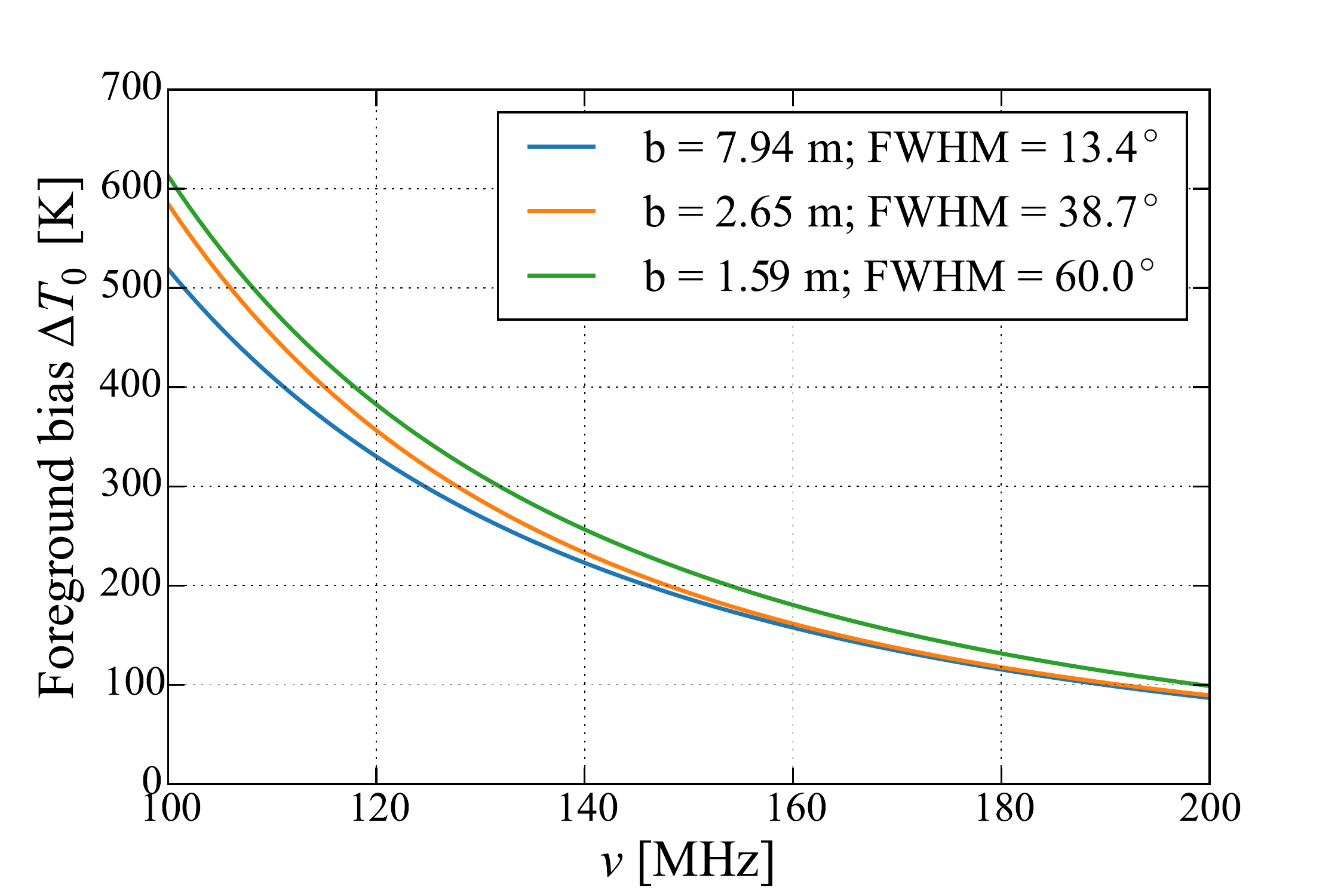}
	\caption{\acl{Changed figure and caption} The foreground bias $\Delta T_0$ as a function of frequency $\nu$ from simulations of an interferometric recovery of the global signal, averaged over $10,000$ random realizations of the simulated foregrounds. The data is shown for three different $6\times6$ square grid arrays of tightly-packed antennae. The blue, orange, and green arrays are composed of antennae with full-width-half-max (FWHM) beam sizes of $13.4^\circ$, $38.7^\circ$, and $60^\circ$ at $100\,\textrm{MHz}$, respectively. Note that narrower beams have less pre-subtraction foreground bias, as expected.}
	\label{fig:unsub_T0_beamSize}
\end{figure}

In Figure \ref{fig:unsub_T0_beamSize} we show simulations of an interferometric recovery of the global signal, $\widehat{T}_0 (\nu)$, averaged over $10,000$ random realizations of the simulated foregrounds. Because our simulations do not contain any cosmological signal, this is equal to $\Delta T_0$, the expected foreground bias. Each curve shows the result for a $6\times6$ square grid \acl{Changed array size} of tightly packed antennas with varying primary beam widths (and therefore varying baseline lengths). We define a ``tightly packed array" as one where the shortest baselines $b_\textrm{short}$ are given by $b_\textrm{short}/ \lambda \sim \sqrt{8 \ln 2} (2 \sqrt{2} \pi \theta_0)^{-1}$. This expression comes from taking physical extent of a Gaussian aperture (which is ultimately just a theoretical construct) to be its full-width-half-max (FWHM) \acl{Updated here}. As expected, arrays with smaller beams/longer baselines exhibit a lower foreground bias, since our observations are centered around the NGP, causing wider beams to pick up more foregrounds from lower galactic latitudes, where they are typically brighter.

\begin{figure}[h]
	\centering
	\includegraphics[width=0.50\textwidth] {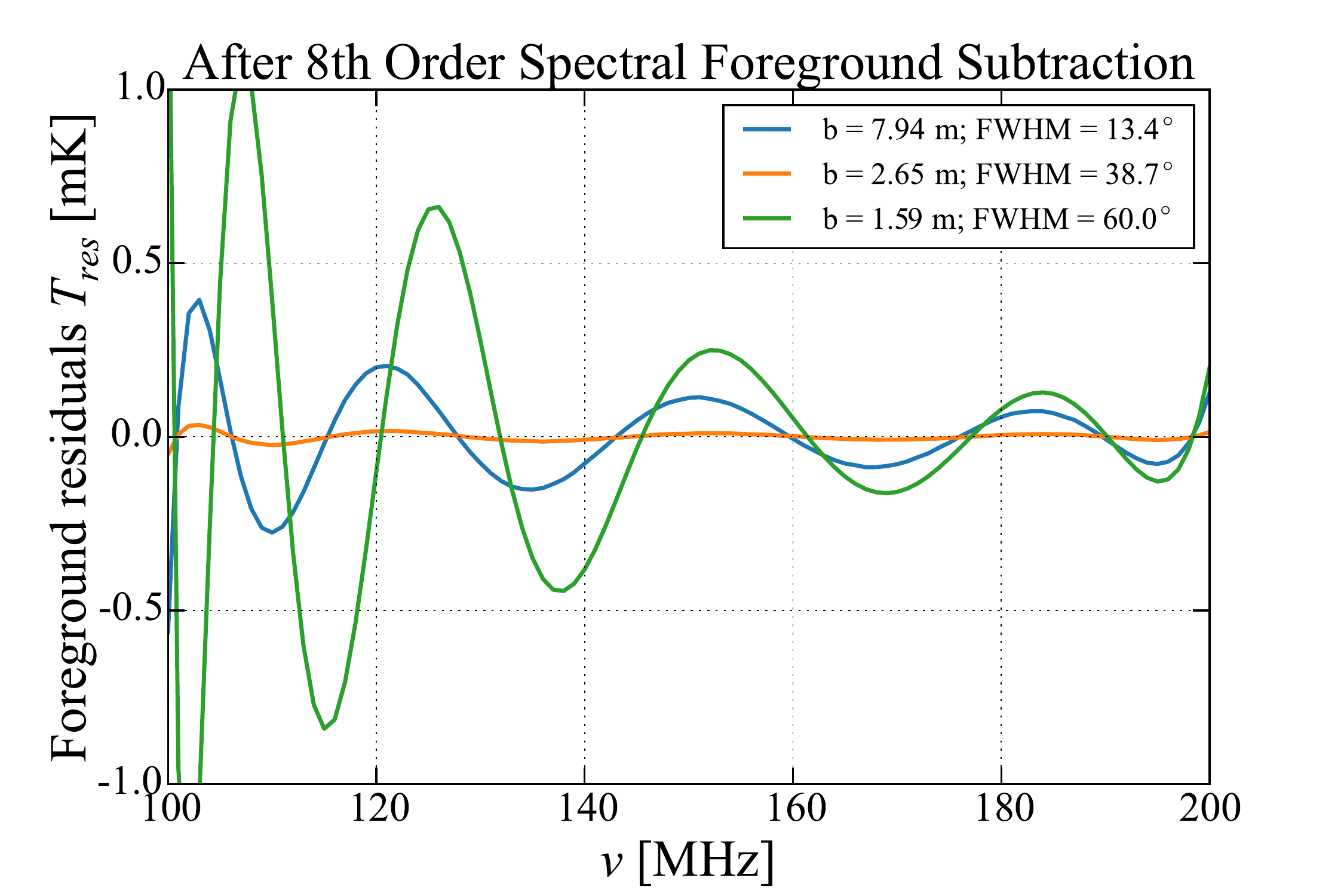}
	\caption{\acl{Changed figure and caption} The foreground residuals after subtracting off a $8$th degree polynomial in log space. The data are shown for the same arrays as from Figure \ref{fig:unsub_T0_beamSize}. Note that unlike in the case of the raw foreground bias, the array with the narrowest beams no longer has the least foreground contamination. Instead, the intermediate-size beam results in the lowest foreground residuals. This is due to the fact that narrower beams come with longer baselines that introduce chromatic ripples that are difficult to subtract.}
	\label{fig:subPoly8_T0_beamSize}
\end{figure}

\begin{figure}[h]
	\centering
	\includegraphics[width=0.50\textwidth] {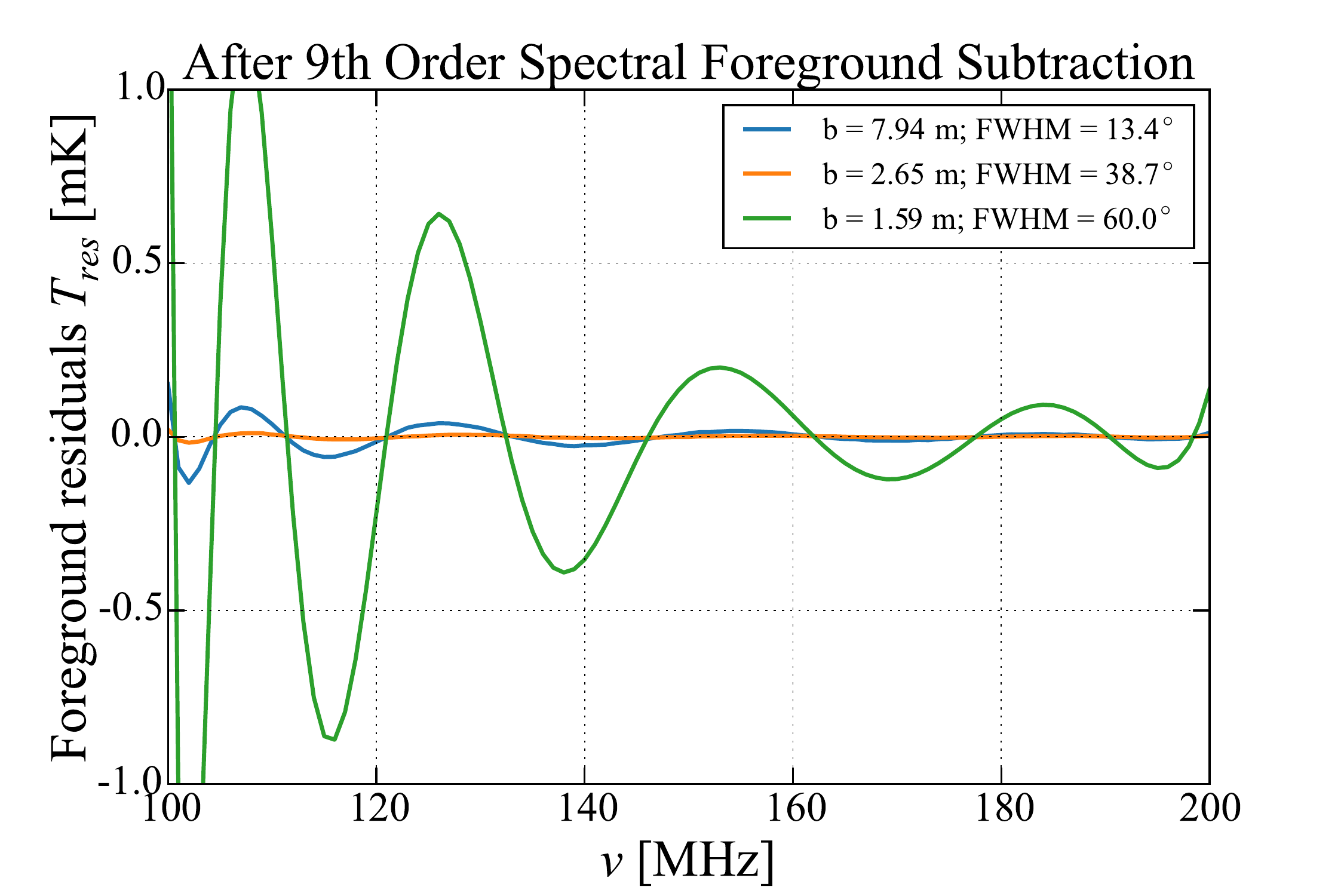}
	\caption{\acl{Changed figure and caption} Same as Figure \ref{fig:subPoly8_T0_beamSize}, except with an $9$th degree polynomial in log space.}
	\label{fig:subPoly9_T0_beamSize}
\end{figure}

\acl{Some trivial updates in the polynomial orders in this paragraph} Figures \ref{fig:subPoly8_T0_beamSize} and \ref{fig:subPoly9_T0_beamSize} show the foreground residuals that remain after the subtraction of $8$th and $9$th order log-space polynomials, respectively. One immediately sees that whereas the narrowest beams/longest baselines gave the dimmest \emph{initial} pre-subtraction spectra, the post-subtraction residuals are minimized for intermediate-sized beams. This is precisely the trade-off that we qualitatively alluded to above: the long baselines that inevitably come with narrow beams cause low-level chromatic ripples in the data that are not easily removed by smooth low-order polynomials, while the broad beams that come with short baselines incorporate brighter lower-latitude foregrounds. Further evidence can be seen by comparing Figures \ref{fig:subPoly8_T0_beamSize} and \ref{fig:subPoly9_T0_beamSize}. One sees that increasing the order of the polynomial fit allows significant further suppression of the chromatic residuals introduced by long baselines, but only results in slight decreases in residuals for the wide beam case. This is because the higher residuals for the latter are the result of an overall increase in foreground amplitude, which affects all polynomial orders. We find the optimal beam size of full-width-half-max $ \sim 40^\circ$ to hold whether considering an experiment spanning the $100$ to $200\,\textrm{MHz}$ band (targeting reionization) or the $50$ to $100\,\textrm{MHz}$ band (targeting the pre-reionization dip), provided the $ 40^\circ$ recommendation refers to the beam size at the lowest observation frequency.

Yet another consideration in choosing baseline length is instrumental noise. As we have already alluded to, short baselines have greater sensitivity to the spatial monopole of the sky, and therefore have better signal-to-noise. Thus, if one were to add instrumental noise to the preceding discussion, the optimal baseline length would shift towards smaller values (with the primary beam correspondingly increasing in width). Picking a baseline length based on the foregrounds-only analysis of this subsection is therefore in principle sub-optimal. In practice, however, instrumental noise is a sub-dominant contribution to the error budgets of most $21\,\textrm{cm}$ global signal experiments, and the optimal baseline length will only be slightly shorter than the one advocated here. Moreover, since interferometers consist of many baselines, the collective signal-to-noise of an array can compensate for a lower signal-to-noise in any individual baseline, a point that we will explore in the following subsection.

\subsection{How many elements?}
\label{sec:numElems}
Thus far, we have established that the ideal global signal interferometer is one comprised of elements with FWHM primary beam widths of $\sim40^\circ$, packed as closely together as possible. However, we have yet to specify the number of elements. In this section we imagine a regular square grid of $N \times N$ antennas and ask what value of $N$ is required for an interferometer to perform as well as a single element in a measurement of the global signal.

As one adds more and more elements to a regular array (increasing $N$), the main effect is an increase in the number of short baselines, providing repeated measurements of the same visibilities that can be combined to average down instrumental noise. While it is true that adding more elements to an array also gives rise to some longer baselines (since the only way to add antenna elements to a closely packed array is to add them to the periphery), these baselines have minimal response to the global signal and only provide information regarding foregrounds. This information can in principle be used to help with foreground mitigation, but as we shall see when we discuss data analysis in Section \ref{sec:MathForm}, it is difficult to use this information without introducing chromatic signatures into the final global spectrum estimates. It is thus safer to severely downweight the influence of long baselines, minimizing their influence on the final result.

With multiple copies of the same baselines, an interferometer can have just as high a signal-to-noise as single element experiment, even if each individual baseline is less sensitive to the global signal. To quantify precisely how many such copies are necessary, we will now compute the expected noise variance from our estimator $\widehat{T}_0 (\nu) $ of the global signal. Starting with Equation \eqref{eq:NormedSimpleEst}, we perturb the $j$th visibility $V(\mathbf{b}_j, \nu)$ by adding an additive noise contribution $n_j (\nu)$. Computing the variance $\boldsymbol \Pi(\nu,\nu)$ of the final estimator then gives
\begin{eqnarray}
\boldsymbol \Pi(\nu,\nu) &\equiv& \langle \widehat{T}_0 (\nu)^2 \rangle - \langle  \widehat{T}_0 (\nu) \rangle^2 \nonumber \\
&=&  \frac{\sigma^2}{ \sum_k \big{|} \int  A(\rhat, \nu) \exp\left(i 2\pi \frac{\nu }{c}\mathbf{b}_k \cdot \rhat \right)d\Omega \big{|}^2},
\end{eqnarray}
where we have assumed that the instrumental noise has zero mean, so that $\langle n_j \rangle = 0$. We have additionally assumed that the noise is uncorrelated between baselines with variance $\sigma^2$, so that $\langle n_i n_j^* \rangle = \sigma^2 \delta_{ij}$. Making the approximation that the sensitivity of the array to the global signal is dominated by the shortest baseline $\mathbf{b}_\textrm{short}$ of which there are $N_\textrm{short}$ copies, the noise variance of an interferometer-estimated global signal reduces to
\begin{equation}
\boldsymbol \Pi(\nu,\nu) \approx \frac{\sigma^2}{ N_\textrm{short} \big{|} \int  A(\rhat, \nu) \exp\left(i 2\pi \frac{\nu }{c}\mathbf{b}_\textrm{short} \cdot \rhat \right)d\Omega \big{|}^2}.
\end{equation}
On the other hand, for a single element experiment we have only a single baseline of length zero, so the noise variance is
\begin{equation}
\boldsymbol \Pi(\nu,\nu) = \frac{2 \sigma^2}{\big{|} \int  A(\rhat, \nu) d\Omega \big{|}^2},
\end{equation}
where the extra factor of $2$ arises from the fact that output voltages are squared in auto-correlation experiments, resulting in a squaring of the Gaussian noise contribution. The variance of the squared noise then depends on the fourth moment of a Gaussian distribution, which gives an extra factor of $2$ when expressed in terms of the variance of the signal, $\sigma^2$. Equating our last two expression allows one to solve for the number of short baselines $N_\textrm{short}$ that are needed for an interferometer to have the same thermal noise sensitivity as a single element experiment:
\begin{equation}
\label{eq:Nshort_curved}
N_\textrm{short} \approx  \frac{\big{|} \int  A(\rhat, \nu) d\Omega \big{|}^2}{  2\big{|} \int  A(\rhat, \nu) \exp\left(i 2\pi \frac{\nu }{c}\mathbf{b}_\textrm{short} \cdot \rhat \right)d\Omega \big{|}^2}.
\end{equation}
To get a rough sense for the magnitude of $N_\textrm{short}$, suppose we make the flat-sky approximation. This gives
\begin{equation}
N_\textrm{short} \approx \frac{\big{|} \widetilde{A}(0)\big{|}^2}{2 \big{|} \widetilde{A}(\nu \mathbf{b}_\textrm{short} / c)\big{|}^2}.
\end{equation}
For Gaussian beams, we obtain
\begin{equation}
N_\textrm{short} \approx \frac{1}{2} \exp \left( 4 \pi^2 \theta_b^2 \frac{b_\textrm{short}^2}{\lambda^2} \right).
\end{equation}
\acl{Edited here for factor of 2. We might also want to talk about how sensitive this is to antenna design.} Here, we see that $N_\textrm{short}$ scales very strongly with baseline length. While this is due
in part to our use of a Gaussian beam, the strong scaling is fundamentally due to the fact that
sensitivity to the global signal drops rather rapidly with increasing baseline length. Again defining a closely packed array to be one with $b_\textrm{short}/ \lambda \sim \sqrt{8 \ln 2} (2 \sqrt{2} \pi \theta_0)^{-1}$, our expression gives $N_\textrm{short} = 8$. For an $N \times N$ square grid of antennas,\footnote{Our translation from $N_\textrm{short}$ to $N$ makes the crucial assumption that independent baselines have independent instrumental noise contributions. Recent calculations (A. Neben, private communication) have suggested that this may not be a good approximation, particularly for tightly packed arrays that are designed to be sensitive to the global signal. However, since the number of antennas required is reasonably small, a straightforward solution to this problem is to simply construct a large number of two-element mini-interferometers rather than a larger single interferometer, with the mini-interferometers placed far away from each other to reduce correlated noise effects.} there are $N_\textrm{short} = 2(N-1)^2$ shortest baselines formed by adjacent antenna pairs (half of which are in one direction, while the other half are perpendicular). Solving for $N$ then gives $N=3$, and therefore even a small array will allow an interferometric measurement to more than make up for the loss of sensitivity to the global signal from discarding auto-correlations.

\acl{Added this paragraph} Admittedly, the calculation that we have just presented is rather sensitive to the details of one's antenna pattern, and the final result of $N=3$ is particular to our Gaussian model. In practice, antennas to be used for a global signal interferometer ought to be carefully designed to ensure that the drop-off in sensitivity to the monopole is not too rapid. Fortunately, existing antenna designs already have a reasonable performance as far as thermal noise is concerned. The $150\,\textrm{MHz}$ PAPER beam model used to generate Figure \ref{fig:effAperture_PAPER}, for example, gives $N_\textrm{short} \sim 32$ (using the full curved sky expression given in Equation \ref{eq:Nshort_curved}) when the $2\,\textrm{m}$-sized PAPER antennas are placed next to each other. This requires a $5 \times 5$ array, larger than our previous estimate but certainly not an unreasonably large number.
%

\section{Data analysis choices}
\label{sec:MathForm}

In the previous section, we examined the trade-offs associated with the design of an interferometer targeting the global $21\,\textrm{cm}$ signal. Qualitatively, we concluded that one ought to design an interferometer array with a modest number of antenna elements, packed as closely together as possible. Ideally, the antenna elements should possess primary beams that are neither too narrow nor too broad, with roughly a full-width-half-maximum of $\sim 40^\circ$ at the lowest observation frequency.

In this section, we assume that an appropriate array has been constructed, and consider instead various trade-offs in data analysis. Inspired by the near-separable models of \citet{Liu_Switzer_2014}, our proposed analysis methods will usually involve a two-step process. First, the data are analyzed frequency-by-frequency, producing an estimate of the spatial monopole at every frequency channel. Following the per-frequency analysis, the data are combined into a single global signal spectrum, where we take advantage of the long frequency-coherence length of foregrounds to perform a final foreground subtraction. In what follows, we will examine the pros and cons of various choices in the detailed implementation of each of these steps.

\subsection{Step 1: Extracting the Spatial Monopole from Visibilities}

We begin with a more general version of Equation \eqref{eq:Vb}, our measurement equation. Discarding the flat-sky approximation, we have
\begin{equation}
V(\mathbf{b}) = \int  T(\mathbf{\hat{r}}) A(\mathbf{\hat{r}}) \exp \left( -i 2 \pi \frac{\nu}{c} \mathbf{b} \cdot \mathbf{\hat{r}} \right) d\Omega,
\end{equation}
where $\mathbf{\hat{r}}$ is a unit vector that specifies locations on the sky. Expressing the sky in terms of spherical harmonics gives
\begin{equation}
\label{eq:VbSphHarm}
V(\mathbf{b}) = \sum_{\ell m} \left(  \int  Y_{\ell m} (\mathbf{\hat{r}}) A(\mathbf{\hat{r}}) e^{ -i 2 \pi \frac{\nu}{c} \mathbf{b} \cdot \mathbf{\hat{r}}} d\Omega \right) a_{\ell m}.
\end{equation}
Since this equation is linear, we may write it as a matrix equation of the form
\begin{equation}
\y = \Q \mathbf{x} + \mathbf{n}
\label{eqn:yQxn}
\end{equation}
where $\y$ is a vector of length $\Nbl$ (the number of baselines) containing the visibilities measured at different baselines $V(\mathbf{b})$, and where we have added an instrumental noise contribution $\mathbf{n}$. The matrix $\Q$ is the beam response of an antenna array at different baselines (rows) and different spherical harmonics (columns). Comparing Equations \eqref{eq:VbSphHarm} and \eqref{eqn:yQxn}, we see that the explicit form of $\Q$ is given by 
\begin{equation}
\textrm{Q}_{j,\ell m} = \int d\Omega A(\mathbf{\hat{r}}) Y_{\ell m}(\mathbf{\hat{r}}) e^{-2\pi i \frac{\mathbf{b_\textit{j}}}{\lambda} \cdot \mathbf{\hat{r}}}
\label{eqn:Qdef}
\end{equation}
where $A$ is the primary beam for the antennas, $\mathbf{b_{\textit{j}}}$ is the $j$th baseline, and $\lambda$ is the wavelength of observation. The sky is represented by $\mathbf{x}$, which is a vector containing all the spherical harmonic coefficients $a_{\ell m}$. As such, it has length $(\ell_{\textrm{max}}+1)^2$, where $\ell_{\textrm{max}}$ is the largest $\ell$ value used in the model of the true sky. The global signal that we seek is proportional\footnote{Throughout this paper, we adopt a spherical harmonic normalization convention where $Y_{00} (\mathbf{\hat{r}}) = 1/ \sqrt{4\pi}$. A pure monopole $T_0$ then has $a_{00} = \int Y_{00} T_0 d\Omega = \sqrt{4\pi} T_0$, so to recover an estimate of $T_0$ from an estimate of $a_{00}$, one must divide by $\sqrt{4 \pi}$.} to the first component of $\mathbf{x}$, i.e., $a_{00}$. Ultimately, then, we only need to form an estimator $\hat{a}_{00}$ of this first component. However, it is crucial to bear in mind that the true $\mathbf{x}$ contains foregrounds with significant power in higher $(\ell, m)$ modes, and that this power may leak into our estimate $\hat{a}_{00}$ of $a_{00}$. The formalism that we present below will provide exactly the right machinery for quantifying such leakage.
%

Since $\Q$ is not in general invertible (or even a square matrix), we cannot solve Equation \eqref{eqn:yQxn} for $\mathbf{x}$ directly and instead can only recover an estimator for $\mathbf{x}$. We consider an estimator of the form 
\begin{equation}
\mathbf{\hat x} = \M \Q^\dagger \Hmat \y,
\label{eqn:xhat}
\end{equation}
where $\Hmat$ and $\M$ are both matrices that can be chosen by the data analyst. The $\Hmat$ matrix is of size $N_\textrm{bl} \times N_\textrm{bl}$, and its role is to encode how the data analyst might wish to weight the different visibilities in forming estimates of the various spherical harmonics. Note that this weighting is in addition to the weighting that is naturally provided by $\Q^\dagger$, which (for a given spherical harmonic mode) upweights baselines if their visibilities are sensitive to that mode and downweights them otherwise. The matrix $\M$ measures $(\ell_\textrm{max} +1)^2 \times (\ell_\textrm{max} +1)^2$, and therefore operates on a set of estimated spherical harmonic modes. Its role is to normalize our estimates of the different modes, and to possibly disentangle them from each other if there is leaked power between the modes. Our choices for $\M$ and $\Hmat$ will determine the statistical properties of our final global signal estimates. For example, the variance $\boldsymbol \Sigma$ of our estimator (the square root of which gives the error bars from instrumental noise) is given by
\begin{equation}
\label{eq:NoiseMatrixSigma}
\boldsymbol \Sigma \equiv \langle \xhat \xhat^\dagger \rangle - \langle \xhat \rangle \langle \xhat \rangle^\dagger = \M \Q^\dagger \mathbf{H} \N \mathbf{H} \Q \M,
\end{equation}
where $\mathbf{N} \equiv \langle \y \y^\dagger \rangle - \langle \y \rangle \langle \y \rangle^\dagger$ is the instrumental noise covariance of the visibilities. This is clearly affected by our choices for $\mathbf{H}$ and $\M$.

\subsection{A possible choice for $\M$}
\label{sec:badMmatrix}
As a first guess, picking
\begin{equation}
\M = [\Q^\dagger \Hmat \Q]^{-1}.
\label{eqn:M}
\end{equation}
might be considered an attractive choice.
The final estimator has the desirable property that its ensemble average $\langle \mathbf{\hat x} \rangle$ satisfies the condition $\langle \mathbf{\hat x} \rangle = \mathbf{x}$. This means that on average, the estimator for a particular spherical harmonic coefficient $\hat{a}_{\ell m}$ is equal to the true coefficient ${a}_{\ell m}$, and there is no leakage between different spherical harmonic modes. An estimate of the monopole is thus truly an estimate of the monopole only, with no contributions from other modes.

However, the $\M$ matrix defined by Equation \eqref{eqn:M} makes an assumption that may not be justified---it assumes that the combination $[\Q^\dagger \Hmat \Q]$ is invertible. Essentially, since the inversion results in $\langle \mathbf{\hat x} \rangle = \mathbf{x}$, we can reverse our line of reasoning to see that any time our observations do not allow different $\hat{a}_{\ell m}$ to be perfectly disentangled from each other, $\Q^\dagger \Hmat \Q$ will be uninvertible. For example, if only part of the sky is surveyed, the best that one can do is to measure linear combinations of different $\hat{a}_{\ell m}$s. Even if the full sky is surveyed (for example by observing the sky with a hypothetical wide-field instrument located at the equator), it is typically difficult to design a broadband instrument that allows for a full inversion without sacrificing the design principles of the previous section, as we will now show.

To perfectly isolate a given spherical harmonic coefficient $a_{\ell m}$ , it is necessary to incorporate information from multiple baselines, each of which measures a slightly different linear combination of spherical harmonics on the sky. A clean extraction requires the data analyst to form yet another linear combination, this time of different visibilities. The goal of this linear combination is to invert the original linear combination of spherical harmonics that was formed by the instrument. Clearly, a necessary condition for this inversion to be successful is for there to be at least as many constraints (i.e., unique visibilities) as there are spherical harmonic coefficients to estimate. Ideally, an array ought to make many independent measurements per spherical harmonic mode to ensure a clean separation of modes.  Since $\ell \sim 2 \pi u$, different $\ell$ modes are separated by $\Delta \ell \sim 2 \pi \Delta u$.  Given that $\ell$ can only take on integer values, this means that having enough measurements is tantamount to requiring our baselines be separated from each other by less than $\Delta u = 1/ 2 \pi$ on the $uv$ plane.  As a concrete example, imagine the square grid of antennas from the previous section, where neighboring antennas separated by a distance $b_\textrm{short}$.  Baselines of this array will also form a square grid of points on the $uv$ plane with the $u$ and $v$ coordinates given by integer multiples of $\Delta u = b_\textrm{short} / \lambda = b_\textrm{short} \nu / c$.  Therefore, in order to have enough measurements for inversion, we must satisfy the condition
\begin{equation}
\label{eq:WantAll}
b_\textrm{short} \ll \frac{c}{2 \pi \nu_\textrm{max}},
\end{equation}
where we have evaluated our constraint at the maximum frequency $\nu_\textrm{max}$ we wish to probe, since that is where it is the most stringent.  On the other hand, as we have argued above, physical constraints on antennas size dictate a spacing satisfying\footnote{Note that strictly speaking, this constraint only applies to the shortest baselines, since it is possible to obtain sub-element sized spacings of longer baselines by slightly dithering the positions of antennas in a large array. For now we will disregard this interesting point because it is the shortest baselines that provide the greatest access to the global signal. However, in future work it may be possible to use sub-element dithering to cleanly measure high $\ell$ modes of the sky, providing extra data-derived information on foregrounds.}
\begin{equation}
\label{eq:AntSize}
b_\textrm{short} \ge \frac{c}{2 \pi \theta_b \nu_\textrm{min}},
\end{equation}
where this time the tightest constraint occurs at the lowest frequency $\nu_\textrm{min}$.


The two constraints listed above make it difficult to probe a large frequency range with a single interferometer.  To see this, note that the upper limit on $b_0$ decreases with increasing $\nu_\textrm{max}$, while the lower limit increases with decreasing $\nu_\textrm{min}$.  With a wide enough frequency range, these two limits meet, and to avoid inconsistent constraints, we require
\begin{equation}
\theta_b \ge \frac{\nu_\textrm{max} }{\nu_\textrm{min}}
\end{equation}
as a \emph{minimum} beam size.  Since this critical beam size depends on the ratio of $\nu_\textrm{max}$ to $\nu_\textrm{min}$, it is easier to satisfy our bounds with a narrowband instrument at higher frequencies, which is precisely the scenario that is uninteresting for a global $21\,\textrm{cm}$ signal experiment. Moreover, since $\nu_\textrm{max}$ must be greater than $\nu_\textrm{min}$ the generosity of our bound saturates at $\nu_\textrm{max} = \nu_\textrm{min}$, and our condition then requires that $\theta_0 \ge 1\,\textrm{rad}$. Recalling that $\theta_b$ is the standard deviation of a Gaussian beam (and not the FWHM), we see that essentially one needs horizon-to-horizon beam coverage of the sky. As discussed previously, however, such a beam would be too wide to be optimal, as it would not allow the selective isolation of cold patches in the galaxy. Indeed, one can see from Figures \ref{fig:subPoly8_T0_beamSize} and \ref{fig:subPoly9_T0_beamSize} that with a FWHM of $60^\circ$, foreground residuals decrease rather slowly.
%

From this, we see that if one is to adhere to the design principles of Section \ref{sec:BackOfEnvelopeArrayDesign}, the conditions required for $\Q^\dagger \Hmat \Q$ to be invertible will necessarily be violated at some observation frequencies. This problem can be alleviated slightly if one is willing to construct multiple narrowband arrays to collectively cover a wide frequency range. However, this is not only an expensive solution, but also a relatively ineffective one---the best that one can do is to pursue an extreme approach where a different array is constructed (or a single array reconfigured) for every observation frequency, but that corresponds precisely to the $\nu_\textrm{max} = \nu_\textrm{min}$ case discussed above, and we have already seen that the required beam sizes are still too wide.

Alternatively, one may simply replace the inverse of $\Q^\dagger \Hmat \Q$ with its pseudo-inverse whenever the matrix $\Q^\dagger \Hmat \Q$ is uninvertible. In doing so, however, one runs the risk of imprinting sharp spectral features into the final estimate of the global signal. This is because a pseudo-inverse inverts only the modes of a matrix that are present with non-zero eigenvalue, and the set of modes that are present will in general be frequency-dependent. The imprint of sharp spectral features should be avoided at all costs, since sharp features have the ability to masquerade as the cosmological signal.

For all the reasons listed above, we therefore recommend against the use of Equation \eqref{eqn:M} for $\M$. 
%

\subsection{A better choice for $\M$}
\label{sec:BetterM}

As an alternative to Equation \eqref{eqn:M}, consider the diagonal matrix given by 
\begin{equation}
\label{eq:diagM}
\M_{ij} = \frac{\delta_{ij}}{(\Q^\dagger \Hmat \Q)_{ii}},
\end{equation}
where $\delta_{ij}$ is the Kronecker delta. With this choice of $\M$, each $a_{\ell m}$ estimate (each component of $\hat{\mathbf{x}}$) is a linear combination of the true $a_{\ell m}$ coefficients. At first sight, this seems to be a drawback of this choice for $\M$, since our goal is to measure the cosmological monopole. However, we will soon use a variety of illustrative special cases to see that this is not so, and that some leakage between different $a_{\ell m}$ coefficients---if appropriately constrained---is in fact a feature. To quantify this leakage, we take the ensemble average of Equation \eqref{eqn:xhat} and insert Equation \eqref{eqn:yQxn}. This yields
\begin{equation}
\langle \xhat \rangle = \M \Q^\dagger \Hmat \Q \x
\end{equation}
Defining a window function matrix $\W$ as 
\begin{equation}
\W \equiv \M \Q^\dagger \Hmat \Q,
\label{eq:Wform}
\end{equation}
we see that $\langle \xhat \rangle  = \W \mathbf{x}$, so each row of $\W$ gives the linear combination of spherical harmonics actually probed by our estimator $\hat{a}_{\ell m}$. The matrix $\W$ quantifies the amount of leakage between modes, and it depends on both our instrument (via $\Q$) and our data analysis method, via Equations \eqref{eqn:xhat} and \eqref{eq:diagM}. Of particular interest will be the first row of $\W$, which will tell us what combination of spherical harmonics are actually being measured when we attempt to constrain the monopole $a_{00}$ mode.

The choice of $\M$ used here has several attractive properties. In Appendix \ref{minVarProof} we prove that if $\Hmat$ is used as an inverse noise covariance weighting of visibilities (i.e., if we have $\Hmat = \N^{-1}$), then our diagonal choice for $\M$ minimizes the variance (and therefore the error bars) on $\mathbf{\hat{x}}$. Additionally, Equation \eqref{eq:diagM} has the property that diagonal elements of $\W$ always equal unity, regardless of what $\Hmat$ is used. Focusing on the first row then, we have $\W_{11} =1$, which implies that the amplitude of a pure monopole sky is preserved by our measurement and data analysis procedures. In other words, there is by construction never any signal loss in this stage of the analysis, where we combine visibilities into spherical harmonic coefficients.
%
%

\subsubsection{Single Element Limit}

Consider the single-element limit as an illustrative example of how our choice of $\M$ works and how leakage between spherical harmonic modes can be a desirable feature. The single-element limit is representative of auto-correlation experiments such as EDGES. With a single antenna element, the matrix $\mathbf{Q}$ reduces to a single row vector consisting of Equation \eqref{eqn:Qdef} evaluated at baseline length $b_j=0$. The measurement vector $\mathbf{y}$ becomes a single measurement $y$, given by the primary beam integrated over the sky:
\begin{equation}
y = \int T(\mathbf{\hat{r}}) A(\mathbf{\hat{r}}) d\Omega.
\end{equation}
Equations \eqref{eqn:xhat} and \eqref{eq:diagM} then reduce to
\begin{equation}
\mathbf{\hat{x}}_{\ell m} = \frac{\int d\Omega Y_{\ell m} (\mathbf{\hat{r}}) A(\mathbf{\hat{r}})}{\left[ \int d\Omega A(\mathbf{\hat{r}}) \right]^2} y.
\end{equation}
For the global signal (i.e., spatial monopole), we are interested in the first component of this $\mathbf{\hat{x}}$ vector. Isolating this and dividing both sides $\sqrt{4\pi}$ to convert from $a_{00}$ to the spatial mean of the sky, we obtain
\begin{equation}
\label{eq:singleElementExtraction}
\widehat{T}_0 = \frac{y}{ \int d\Omega A(\mathbf{\hat{r}}) },
\end{equation}
which is the estimator that one would have guessed from simple considerations---the measurement $y$ is just a weighted average of the sky, and the denominator normalizes the weights. This is in fact also a special case of the estimator given in Equation \eqref{eq:singleBlSillyEst}, where the baseline vector $\mathbf{b}$ is set to zero. Note that there was no need to specify the matrix $\Hmat$ because, with only a single measurement from a single element, $\Hmat$ reduces to a single scalar. The copy of $\Hmat$ in Equation \eqref{eqn:xhat} will therefore always cancel the copy in Equation \eqref{eq:diagM}.
%

Explicitly evaluating Equation \eqref{eq:Wform} for our single element case, the window function for the sky monopole (i.e., the first row of $\W$) is given by
\begin{equation}
\W_{0}(\ell,m) = \frac{\int d\Omega A(\mathbf{\hat{r}}) Y_{\ell m} (\mathbf{\hat{r}})}{\int d\Omega A(\mathbf{\hat{r}}) / \sqrt{4 \pi}}.
\label{eq:singleDipoleW0}
\end{equation}
Naively, one might have hoped for $W_0$ to be zero for all values of $\ell$ and $m$ except for $(\ell, m) = (0,0)$, so that our estimator for the sky monopole does not contain any leaked power from other spherical harmonics. This is what the $\M$ matrix of Section \ref{sec:badMmatrix} would have achieved and is what was sacrificed by our new choice of $\M$. On closer examination, however, the leakage appears to be rather innocuous and is simply the result of our having only surveyed a small portion of the sky (the part that lies within the primary beam). Confirming this interpretation is the form of the numerator in Equation \eqref{eq:singleDipoleW0}, which is the spherical harmonic decomposition of the primary beam.

As discussed above in Section \ref{sec:BackOfEnvelopeArrayDesign}, it is advantageous to concentrate observations on cooler parts of the sky. The non-zero width of the window function given by Equation \eqref{eq:singleDipoleW0} is thus a desirable feature. Another way to see this is to recognize that focusing on a small patch of the sky (and therefore accepting our broader window function) gives us a better estimate of the \emph{cosmological} monopole signal than if we were somehow able to force the window function to be zero away from $(\ell,m) = (0,0)$. In the latter case we would be better estimating the monopole signal of total sky emission, but much of this would be due to the foreground contribution. The key point is that the foregrounds also have a monopole, and without an \emph{a priori} way to distinguish between the monopole of the cosmological and the monopole of the foregrounds, making a ``clean" measurement of the sky's monopole simply adds stronger foregrounds.

\subsubsection{Interferometric case}
\label{sec:MultiBaselineWorkedExample}
We now turn to the interferometric, multi-baseline case. With our measurement $\mathbf{y}$ now consisting of more than just a single number, there is the opportunity to weight our data in non-trivial ways. Put another way, it will be necessary to decide on a form for $\Hmat$, since its two copies will no longer cancel when forming $\xhat$.

Consider an inverse noise covariance weighting of $\Hmat = \mathbf{N}^{-1}$. With the assumption that instrumental noise is uncorrelated and uniform across different baselines, this is equivalent to $\Hmat = \mathbf{I}$. The window function matrix then simplifies to $\W \propto \M \Q^\dagger \Q$, with the key piece being $\Q^\dagger \Q$. The $\M$ matrix is irrelevant to any discussion of leakage between different spherical harmonics, since the form given by Equation \eqref{eq:diagM} is diagonal, and thus the matrix only provides a normalization for each spherical harmonic without further mixing between modes. Evaluating the window function matrix explicitly, we obtain
\begin{eqnarray}
\W_{\ell m, \ell^\prime m^\prime} &\propto& \left( \Q^\dagger \Q \right)_{\ell m, \ell^\prime m^\prime} \nonumber \\
& \propto & \sum_k \left( \int d\Omega A(\rhat) Y^*_{\ell m}(\rhat) e^{i 2 \pi \frac{\mathbf{b}_k}{\lambda} \cdot \rhat }\right) \nonumber \\
&& \times \left( \int d\Omega^\prime A(\rhat^\prime) Y_{\ell^\prime m^\prime}(\rhat) e^{-i 2 \pi \frac{\mathbf{b}_k}{\lambda} \cdot \rhat^\prime }\right).
\end{eqnarray}
For the purposes of measuring the global signal, it is again the first row of this matrix that is the most relevant. Making the flat-sky approximation for the sake of intuition yields
\begin{equation}
\label{eq:AnotherApproxWindow}
\W_{0}(\mathbf{u}) \propto \sum_k \widetilde{A}^* \left( \frac{\mathbf{b}_k}{\lambda} \right) \widetilde{A} \left( \mathbf{u} - \frac{\mathbf{b}_k}{\lambda} \right).
\end{equation}
Now, $\widetilde{A}$ is a function that peaks at the origin and drops off on a characteristic scale $(2 \pi \theta_b)^{-1}$. Thus, this expression tells us that the highest $u = | \mathbf{u} |$ scale that is probed by our global signal interferometer is
\begin{equation}
u \sim \frac{b_\textrm{max}}{\lambda} + \frac{1}{2 \pi \theta_b} = \frac{N\sqrt{2} +1}{2 \pi \theta_b},
\end{equation}
where $b_\textrm{max}$ is the longest baseline in the array, and in the last equality we assumed a closely packed $N \times N$ square array, just as we did in Section \ref{sec:numElems}. The reciprocal of this expression gives the finest angular scale $\theta_\textrm{fine}$ that our interferometer is sensitive to:
\begin{equation}
\label{eq:AngularScaleProbed}
\theta_\textrm{fine} \sim 13.5^\circ \left( \frac{\textrm{FWHM}}{40^\circ} \right) \left( \frac{N \sqrt{2} + 1}{5 \sqrt{2} + 1} \right)^{-1}.
\end{equation}
Here, we have eliminated $\theta_b$ (which corresponds to the standard deviation for a Gaussian beam) in favor of the FWHM, and have used a fiducial array size of $N=5$ to be slightly on the conservative side of our optimal $N=2$ or greater from Section \ref{sec:numElems}. We may thus conclude that an interferometer that is designed in accordance with the principles laid out in Section \ref{sec:BackOfEnvelopeArrayDesign} will not be sensitive to scales finer than $\sim 10^\circ$. Since the anisotropies of the cosmological $21\,\textrm{cm}$ signal are negligible beyond an angular scale of $\sim 1^\circ$ to $2^\circ$ \citep{BittnerLoeb2011}, the modes that are measured by our interferometer are essentially global signal modes, even if they are not formally the $\mathbf{u} = 0$ (or $\ell = m = 0$) mode. Indeed, this argument is one that is implicitly invoked by most theoretical simulations of the global signal---since full sky simulations are too computationally expensive to perform, most (if not all) simulations simply average over an angular field of view that is much greater than the angular scale of anisotropies, and declare the result the global signal.

In short, the leakage of higher spherical harmonic modes into our estimate of the global signal is likely not a concern. In fact, our estimate is a conservative one, because we assumed that the longest baseline of an array contributes significantly to the estimator of the monopole mode. In practice, long baselines have so little response to the monopole that its contribution to our estimator is heavily downweighted by the presence of $\mathbf{Q}^\dagger$. This manifests itself as the $\widetilde{A}^* ( \mathbf{b}_k / \lambda )$ term in our approximate window function, Equation \eqref{eq:AnotherApproxWindow}. The window function for the monopole is therefore preferentially dominated by the baselines that probe broad angular scales.

Importantly, however, our argument relied on the fall-off of $\widetilde{A}$. If the primary beam contains fine features, $\widetilde{A}$ becomes a rather broad function, and one's interferometer begins to be sensitive to the more substantial small-scale modes of the $21\,\textrm{cm}$ anisotropies. This may be an important effect for the global signal measurements performed at LOFAR using lunar occultations, which imprints small spatial structures in the beam \citep{VedanthamLOFAR2}. Fortunately, our formalism provides an easy way to compute the relevant window functions to assess the viability of occultation measurements.

\subsection{Choices for $\Hmat$}
\label{sec:Hchoices}
Having motivated our diagonal choice for $\M$, we now consider our choice for $\Hmat$, which weights the visibilities. We will find that although the presence of $\Hmat$ in our estimator provides additional flexibility that can in principle be harnessed to better suppress foregrounds, a ``simple is best" approach of setting $\Hmat = \mathbf{I}$ is more robust and gives better final results.

To see how $\Hmat$ can aid in foreground mitigation, imagine we were able to accurately model the full covariance matrix of foregrounds $\mathbf{N}_\textrm{fg}$. Picking $\Hmat = \Nfg^{-1}$ in Equation \eqref{eqn:xhat} would then downweight select modes in the visibilities $\y$ that are particularly foreground-contaminated, before the $\Q^\dagger$ matrix converts the collection of visibilities into estimates of spherical harmonic estimates (albeit unnormalized ones, prior to the action of $\M$). In practice, one does not possess sufficient information to construct a full covariance matrix, and approximations must be made.

\subsubsection{Approximating $\Hmat  = \Nfg^{-1}$ as a diagonal matrix in harmonic space}
Suppose that one were to approximate the foreground covariance matrix as diagonal in some judiciously chosen basis. Of course, any matrix is by construction diagonal in its own eigenbasis. However, moving into this basis would require knowing the exact covariance matrix to begin with. Instead, our goal should be to find some basis that sufficiently captures the features of the matrix that are needed for foreground mitigation. Consider, for example, a matrix that is diagonal in harmonic space. Modeling the foreground sky in such a way is tantamount to saying that the foregrounds are statistically isotropic, and therefore describable using a power spectrum. While this may be sufficient for some applications, in our case it is relatively unhelpful. To see this, consider the action of an interferometer in the flat-sky approximation.  Disregarding the primary beam for a moment, each baseline would simply measure a different Fourier mode of the sky. In the full formalism, $\Q$ maps spherical harmonics to visibilities; in the flat-sky approximation, the spherical harmonics and visibilities both reduce to Fourier modes, so $\Q$ correspondingly reduces to $\mathbf{I}$. The $\M$ matrix in our estimator therefore simplifies to $\Hmat^{-1}$, which then acts directly on our $\Hmat$ weighting of the visibilities (since $\Q^\dagger$ is now the identity). The two copies of $\Hmat$ then cancel each other, and the estimator becomes $\xhat = \y$. Conceptually, the visibilities are already a measurement of the harmonics of the sky, and thus the different visibilities never have to mix with each other to produce our final (harmonic) estimator. Any downweighting of a strong foreground mode in harmonic space is then simply upweighted back to its original strength, and no foreground mitigation happens. Of course, in a realistic situation we violate the assumptions of the flat-sky and a uniform primary beam, but the foreground suppression effects are still likely to be minimal.

\subsubsection{Approximating $\Hmat  = \Nfg^{-1}$ as a diagonal matrix in image space versus setting $\Hmat  = \mathbf{I}$}
In contrast, consider a foreground covariance matrix that is diagonal---but not the identity---in image space. When transformed into visibility space, the $\Nfg$ matrix then contains off-diagonal elements. In acting on $\y$ through $\Hmat = \Nfg^{-1}$, different visibilities are then mixed together in an effort to suppress foregrounds. Physically, this corresponds to the statement that if the foregrounds are not statistically isotropic (so that the diagonal elements of the covariance vary), the harmonic coefficients of the sky will end up possessing correlated phases. Since the visibilities are approximately measurements of the harmonic sky, they too will be correlated. These correlations can then be used to downweight the contributions of brighter parts of the foreground sky.

Concretely, suppose we form an image-space covariance matrix $\R$, which is given by
\begin{equation}
\label{eq:Rmatrix}
\R_{ij} = m^2(\rhat_i) \delta_{ij}
\end{equation}
where $m^2(\rhat_i)$ is the model for the foreground sky brightness at the $i$th pixel with angular position $\rhat_i$. To use this in our estimator requires translating this matrix into visibility space by constructing
\begin{equation}
\label{eq:GRG}
\Nfg = \mathbf{G} \R \mathbf{G}^\dagger,
\end{equation}
where
\begin{equation}
\mathbf{G}_{ij} = A(\rhat_j)\exp\left(-i2\pi \frac{\mathbf{b_\textit{i}}}{\lambda} \cdot \boldsymbol \rhat_j\right) \Delta \Omega,
\end{equation}
with $\Delta \Omega$ being the solid angle encompassed by each pixel of our sky model. Note that in general $\Nfg$ is a smaller matrix than $\R$, since the former measures $N_\textrm{bl} \times N_\textrm{bl}$ while the latter measures $N_\textrm{pix} \times N_\textrm{pix}$, where $N_\textrm{pix}$ is the number of pixels in our model.

\begin{figure}[h]
	\centering
	\includegraphics[width=0.50\textwidth] {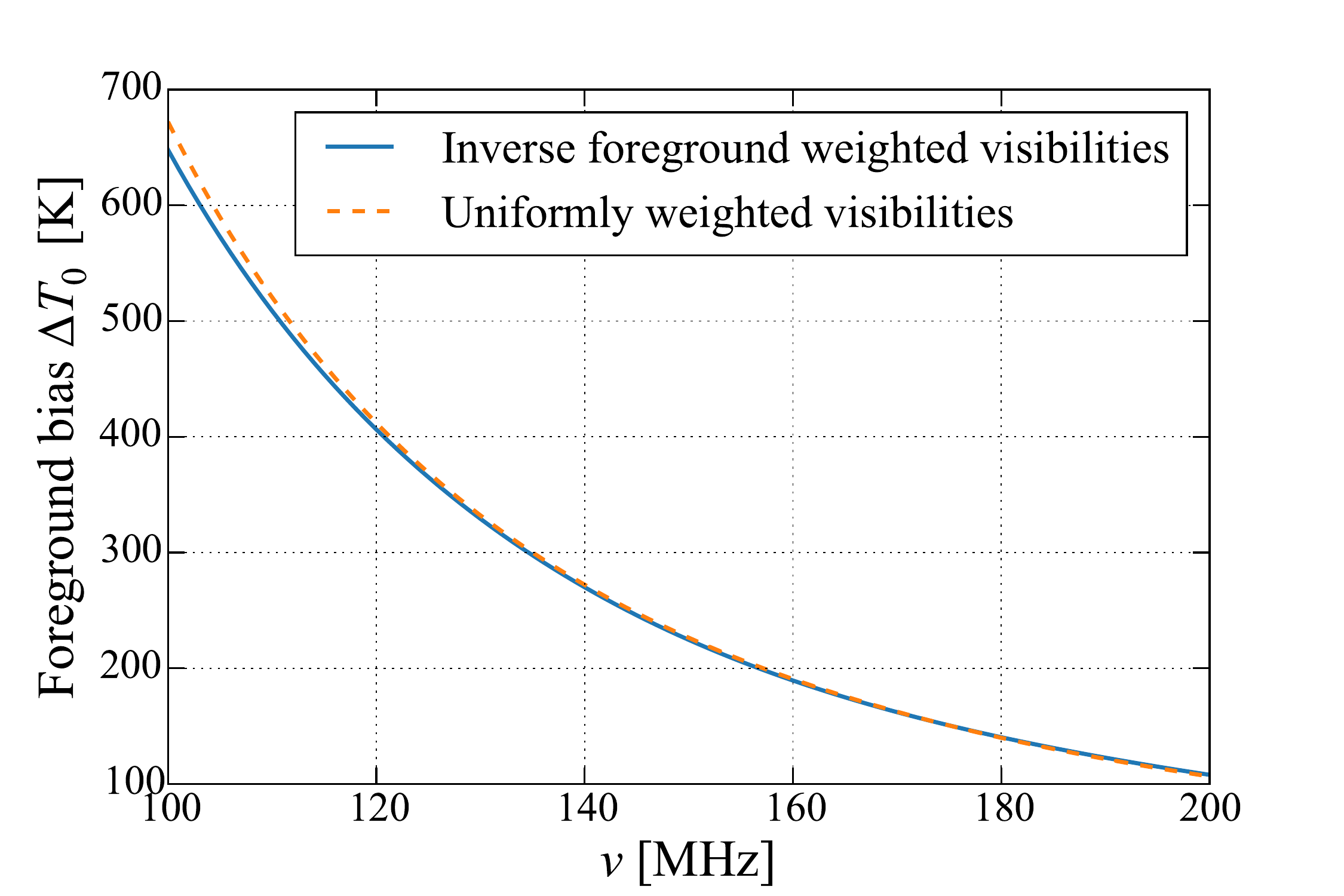}
	\caption{\acl{Updated figure and caption} The foreground bias $\Delta T_0$ as a function of frequency, comparing the data analysis method that weights visibilities by the inverse of the foreground amplitude (the solid blue line) with one that weights visibilities uniformly (the dashed orange line). The experimental setup consists of a closely-packed $6\times 6$ array of antennas with a primary beam of $88^\circ$, with observations centered on the NGP. Adding the foreground-motivated weighting does reduce the foreground bias, but not in any significant way.}
	\label{fig:Ncomparison}
\end{figure}

In Figure \ref{fig:Ncomparison}, we compare the global signal estimators that result from analyzing simulated data using $\Hmat = \Nfg^{-1}$ to those that are obtained with $\Hmat = \mathbf{I}$. For the foreground model in Equation \eqref{eq:Rmatrix}, we use the Haslam map at $408\,\textrm{MHz}$. Note that even though the Haslam map will have the wrong amplitude for observations in our frequency band, the overall amplitude of our foreground map will always cancel out in our final estimator, since $\Hmat$ appears both in Equation \eqref{eqn:xhat} and in our normalization $\M$. In other words, only the angular \emph{shape} of the foreground sky matters. The instrumental simulations used for Figure \ref{fig:Ncomparison} are identical to those of Section \ref{sec:beamSize}, except with the FWHM of the primary beams set to $88^\circ$. With a broader beam, one exaggerates the effect of setting $\Hmat = \Nfg^{-1}$. This is because a larger field of view captures more of the strongly anisotropic nature of foregrounds, which makes a selective downweighting of the brighter parts of the sky more important. Figure \ref{fig:Ncomparison} shows that such a downweighting does in fact reduce the brightness of the foregrounds, but only by a small amount. Part of this is because of the small size of our array, which limits the number of long baselines that are available for resolving fine spatial structures on the sky. However, incorporating longer baselines into our measurement quickly runs afoul of the constraint imposed by Equation \eqref{eq:AngularScaleProbed}.

The reduction in foreground contamination is more pronounced if observations are centered on the galactic plane. There, the foregrounds vary strongly with galactic latitude, making downweightings much more important, and reductions of up to a factor of $\sim 2$ are possible. In practice, however, one tends to avoid observing in the galactic plane anyway. Therefore, the benefits of $\Hmat = \Nfg^{-1}$ are likely to be minimal, particularly when one moves back to using the narrower (but not too narrow) beams suggested in Section \ref{sec:beamSize}, which can more easily isolate patches of the sky that are more approximately isotropic.

\begin{figure}[h]
	\centering
	\includegraphics[width=0.50\textwidth] {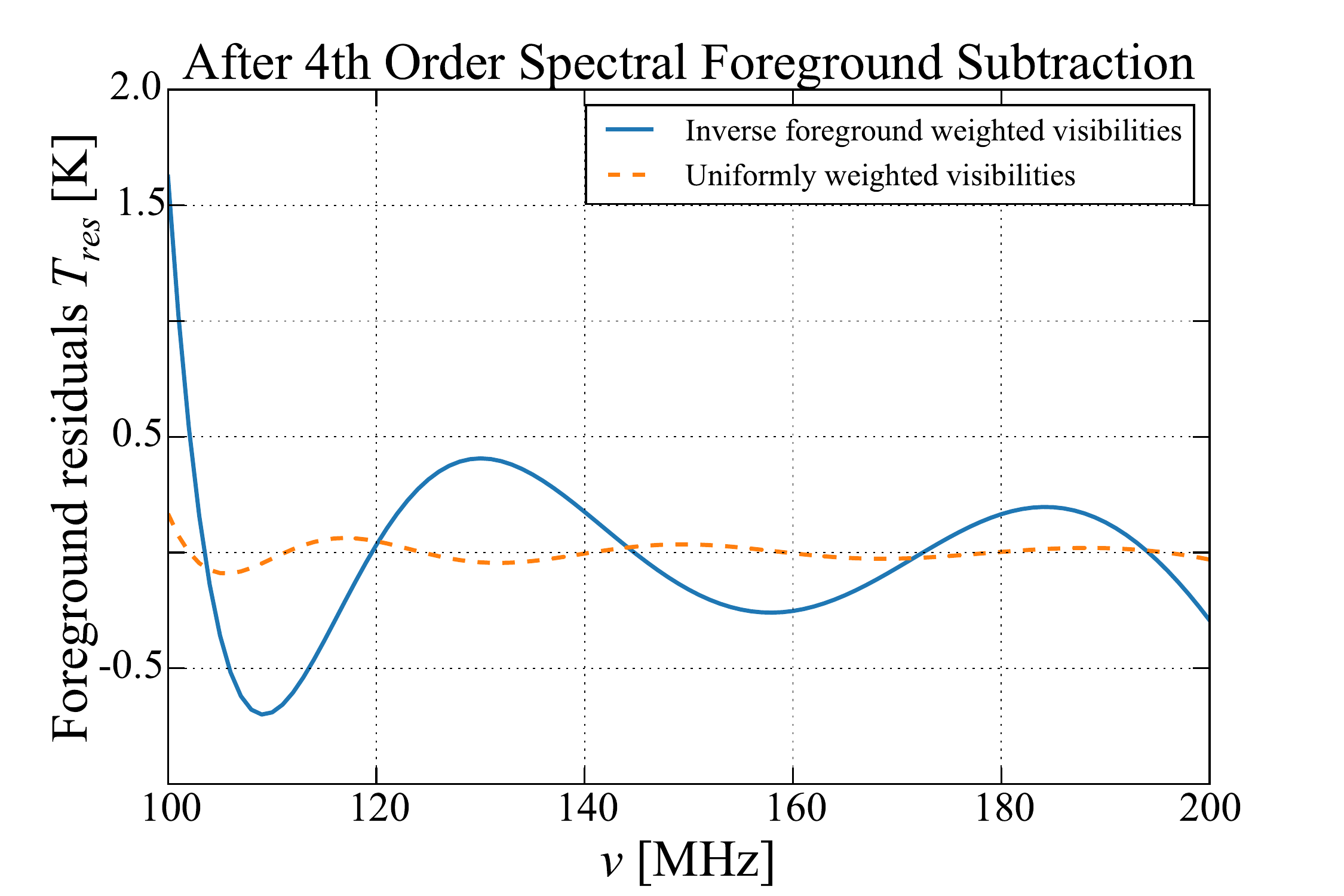}
	\caption{\acl{Updated figure and caption} Foreground residuals from Figure \ref{fig:Ncomparison} after a $4$th order logarithmic polynomial subtraction. The uniformly weighted residuals are now seen to be weaker than those of the inverse foreground weighted analysis. Inverse foreground weighting is thus seen to imprint extra chromaticity in measured spectra, making spectral foreground subtraction more difficult.}
	\label{fig:NcomparisonPoly4}
\end{figure}

\acl{Modified this paragraph to soften some claims} More seriously, setting $\Hmat = \Nfg^{-1}$ has the potential to introduce harmful artifacts into the final spectra. As one begins to subtract smooth components from the spectra, the residuals from the uniformly weighted analysis often become smaller than those from the foreground-weighted analysis. This can be seen in Figure \ref{fig:NcomparisonPoly4}, where we show the residuals after a $4$th order log-polynomial subtraction. This is the result of the foreground-weighted analysis introducing unsmooth structures into the intrinsically smooth spectrum. To be fair, it is possible to find combinations of polynomial orders and instrumental parameters where the foreground-weighted method outperforms uniform weighting. However, it is certainly not uncommon for $\Hmat = \Nfg^{-1}$ to perform worse. Much of this is because the quest to downweight brighter portions of the foreground sky requires the isolation of spatially small patches. To isolate these patches, high angular resolution is necessary, which means that the estimator must weight longer---and more chromatic---baselines more heavily. Long baselines also have the disadvantage of having low sensitivity to the monopole. With a heavier weighting of long baselines, it is more difficult for an interferometer to match the thermal noise sensitivity of a single element experiment. Although in future work it may be possible to eliminate all of these issues with a more sophisticated foreground-motivated form of $\Nfg$, for now we propose the use of $\Hmat = \mathbf{I}$ to be conservative.

\subsection{Step 2: Fitting smooth foregrounds}
\label{sec:fitting}
After the frequency-by-frequency combination of visibilities into an initial estimate of the monopole mode, one obtains spectra such as those shown in Figure \ref{fig:unsub_T0_beamSize} and Figure \ref{fig:Ncomparison}. The spectra are clearly still dominated by smooth, near-power law foregrounds. It is therefore necessary to take further steps to mitigate foregrounds once the data have been reduced to a single spectrum.

One approach is to subtract off smooth functions from the spectrum, be they polynomials or more data-driven forms (such as principal component spectra used in \citealt{Liu_Switzer_2014}). The (hopefully foreground-free) residuals can then be compared to theoretical models for the global signal, although care must be taken to properly account for the possibility that part of the cosmological signal may have been subtracted along with the foregrounds. An alternate approach, which we adopt in the rest of this paper, is to follow in the footsteps of \citet{PritchardLoeb2010,DAREMCMC,BernardiLEDA} and fit for foreground and cosmological model parameters simultaneously. Doing so provides a natural description for signal loss, which manifests itself as degeneracies between foreground parameters and cosmological model parameters. One important trade-off is to decide how many foreground parameters to include. If too many parameters are used, much of the cosmological signal will be absorbed into the foreground model, resulting in large degeneracies and large final error bars on the parameters. On the other hand, having too few parameters will result in foreground residuals that will bias cosmological parameter values. We take the same approach as \citet{BernardiLEDA}, where we include just enough foreground parameters for the cosmological parameter bias to be subdominant to the errors.

\subsection{Summary of data analysis methods}

In summary, we propose a ``simple is best" approach for extracting the global signal from interferometric data. In what follows, we will set $\Hmat = \mathbf{I}$ and adopt Equation \eqref{eq:diagM} for $\M$. Plugging these into Equation \eqref{eqn:xhat}, recasting our vector/matrix expressions in terms of continuous functions, and once again dividing by $\sqrt{4 \pi}$ to convert from an estimate of $a_{00}$ to the global signal, we obtain
\begin{equation}
\label{eq:IntegralEstInterferometer}
\widehat{T}_0 (\nu)= \frac{\sum_j \left[ \int d\Omega A(\rhat, \nu) \exp\left( i 2 \pi \frac{\nu }{c} \mathbf{b}_j \cdot \rhat \right)\right] V(\mathbf{b}_j, \nu)}{\sum_k \big{|} \int d\Omega A(\rhat, \nu) \exp\left( i 2 \pi \frac{\nu }{c} \mathbf{b}_k \cdot \rhat \right) \big{|}^2},
\end{equation}
where we have re-introduced the frequency dependence of various quantities in our notation. Essentially, our estimator amounts to performing a linear fit (frequency-by-frequency) to our data in order to find the value of the monopole that is most consistent with our measured visibility. Following this, we fit the spectrum with a model that includes foreground fits and cosmological parameters. For our foreground fits, we use the same parametric forms as we did in Section \ref{sec:beamSize}, namely Legendre polynomials in $\log \widehat{T}$-$\log \nu$.

\section{Numerical simulations}
\label{sec:SimResults}

In this section, we bring together the various lessons that we have learned regarding instrument design and data analysis to numerically forecast the performance of a fiducial global signal interferometer. Much of our simulation methodology has already been employed in previous sections to produce intermediate results, but we will provide a quick summary here (and add new details) for the reader's convenience.

Guided by the rough arguments of Section \ref{sec:BackOfEnvelopeArrayDesign}, we simulate visibilities from a $6\times6$ \acl{Changed here} square grid of tightly packed antennas. The primary beam of each element is taken to be a tapered Gaussian of the form given by Equation \eqref{eq:TaperedGauss}, with $\theta_b = 0.3\,\textrm{rad}$ (for a FWHM of $38.7^\circ$) at the lowest observation frequency. At higher frequencies, the beam width is assumed to be proportional to $\lambda$. We assume that observations are centered on the Northern Galactic Pole and span a band consisting of $1\,\textrm{MHz}$ channels from $100\,\textrm{MHz}$ to $200\,\textrm{MHz}$ for the reionization, and separately from $50$ to $100\,\textrm{MHz}$ for the pre-reionization epoch. (In other words, we are considering two separate experiments, both with a primary beam FWHM of $38.7^\circ$ at the lowest frequency part of their band). For comparison, we also predict the performance of a single-element global signal experiment using the same type of antenna element and the same observing strategy.

For our simulated foreground sky, we use the same set-up as we did in Section \ref{sec:beamSize}, where each pixel of the $408\,\textrm{MHz}$ map is extrapolated to the relevant frequencies on a pixel-by-pixel basis using Equation \eqref{eq:HaslamExtrap}. The parameters in the power-law-like extrapolation are drawn randomly as before, and we generate $10,000$ different realizations of the foreground sky. With each sky, we then simulate visibilities and total power measurements for the interferometric and single element measurements, respectively. Frequency-by-frequency estimates of the global signal are then obtained using Equation \eqref{eq:singleElementExtraction} for the single element and Equation \eqref{eq:IntegralEstInterferometer} for the interferometer. The results are then averaged together to yield mean foreground spectra for each type of experiment. Finally, smooth foreground components are fit from these spectra\footnote{In principle, one ought to perform foreground fits prior to ensemble averaging the different sky realizations. However, since each line of sight is generated independently in our simulations, ensemble averaging essentially amounts to generating more lines of sight. This in fact results in a more conservative foreground model. To see this, consider a toy example where each line of sight is a randomly drawn power law. Fitting each pixel individually with a power law would be guaranteed to return no residuals. However, since the sum of power laws is not itself a power law, the averaged spectrum over all pixels contains greater curvature, which in general will not be well-fit by a power law. We may therefore safely ensemble average prior to foreground fitting, knowing that the result will be a more conservative foreground spectrum.} in the manner described in Sections \ref{sec:beamSize} and \ref{sec:fitting}. The result is a set of residual foreground spectra.

Aside from residual foregrounds, our forecasts must also incorporate instrumental noise. Modeling this contribution requires three separate covariance matrices. The first is the instrumental noise covariance $\N$ of the visibilities. We assume that the instrumental noise is uncorrelated between different baselines, so that
\begin{equation}
\N_{ij} (\nu) = \frac{T^2_\textrm{sys}(\nu) \Omega_p^2}{t_\textrm{int} \Delta \nu} \delta_{ij},
\end{equation}
where the indices refer to different baselines, $\Omega_p = \int  A(\mathbf{\hat{r}}) d\Omega$ is the size of the primary beam, $T_\textrm{sys}$ is the system temperature, $\Delta \nu$ is the channel width, and $t_\textrm{int}$ is the total integration time. We take $\Delta \nu = 1\,\textrm{MHz}$ and $t_\textrm{int} = 300\,\textrm{hrs}$. For the system temperature, we assume that the measurements are sky-noise dominated, and we set $T_\textrm{sys}$ equal to the primary beam averaged sky temperature.

With the noise covariance of the visibilities $\N$ in hand, we can obtain the covariance matrix $\boldsymbol \Sigma$ of our estimator $\xhat$. To do so, we insert $\N$ into Equation \eqref{eq:NoiseMatrixSigma}. Since $\xhat$ contains estimates of all the spherical harmonic modes that we wish to solve for,   it is an $(\ell_\textrm{max} +1)^2 \times (\ell_\textrm{max}+1)^2$ matrix relating all the errors on the $a_{\ell m}$ estimates to one another. With our focus being the monopole term, we require only the first element on the diagonal of $\boldsymbol \Sigma$. Extracting this element and dividing by $4\pi$ (the square of the conversion between $\hat{a}_{00}$ and $\widehat{T}_0$) gives the variance on $\widehat{T}_0$. Repeating this process for every observation frequency, we can place the resulting variances along the diagonal of yet another covariance matrix $\boldsymbol \Pi$. This is the frequency-frequency noise covariance matrix of our final spectrum, and by populating only its diagonal elements (setting all other elements to zero), we are assuming that noise contributions from different frequencies are uncorrelated. Note that even though we established this procedure for computing $\boldsymbol \Pi$ with interferometers in mind, it can be easily adapted for the single-element experiments as well. Considering a single baseline of length zero, one obtains  $T^2_\textrm{sys}(\nu) \Omega_p / t_\textrm{int} \Delta \nu$ along the diagonal of $\boldsymbol \Pi$, which simply needs to be enhanced by a factor of $2$ to account for the correlated noise discussed in Section \ref{sec:numElems}.

To evaluate the effectiveness of our global signal interferometer, we employ a Fisher matrix formalism. Our set-up is essentially identical to that of \citet{PritchardLoeb2010} and \citet{BernardiLEDA}, \mep{added citation} and thus we relegate a review of the formalism to Appendix \ref{fisher}. As a toy model for the dark ages, we again follow \citet{BernardiLEDA} and model the dip at $\sim 70\,\textrm{MHz}$ as a Gaussian:
\begin{equation}
\label{eq:Dip}
T_\textrm{dip}(\nu) = -A \textrm{exp}\left ( -\frac{(\nu - \nu_0)^2}{2\sigma^2} \right ),
\end{equation}
where $A$ is the amplitude of the signal, $\nu_0$ is the center of the pre-reionization absorption dip, and $\sigma$ is the width. For the reionization signal, we use the form
\begin{equation}
\label{eq:Step}
T_\textrm{reion}(\nu) = \frac{T_{21}}{2} \sqrt{\frac{1+z}{10}}\left[ 1 +  \tanh \left( \frac{z-z_r}{\Delta z} \right)\right],
\end{equation}
where $z_r$ is the redshift of the mid-point of reionization, $\Delta z$ is its rough duration, $T_{21}$ is an overall amplitude, and $z = (1420 \,\textrm{MHz} / \nu) - 1$.

\begin{table*}[htbp]
   \centering
   \begin{tabular}{@{} llcccccc @{}} 
      \toprule
      & & \multicolumn{3}{c}{Reionization} & \multicolumn{3}{c}{Pre-reionization Dip} \\
      \cmidrule(lr){3-5} 
      \cmidrule(lr){6-8} 
      Model &  & $T_{21}$ [K] & $z_r$ & $\Delta z$ & $A$ [K] & $\nu_0$ [MHz] & $\sigma$ [MHz]\\
      \midrule
      Pessimistic & Fiducial Value & 0.010 & 12 & 3 & 0.01 & 60 & 10 \\
      \cmidrule(l){2-8} 
      			 & Interferometer Bias & $2.85\times 10^{-1}$ & $3.07$ & $3.94 \times 10^{-2}$ & $-5.59\times 10^{-1}$ & $-1.77\times 10^2$ & $8.68 \times 10^1$ \\
			 & Single Element Bias & $2.50\times 10^{-1}$ & $7.66\times 10^{-1}$  & $-3.68\times 10^{-1}$  & $9.2\times 10^{-2}$ & $3.07 \times 10^{1}$  & $-1.41 \times 10^1$ \\	 
      			 & Interferometer Error & $\pm 1.11 \times 10^{-1}$ & $\pm 2.97$ & $\pm 2.99$ & $\pm 5.45 \times 10^{-1}$ & $\pm 1.72 \times 10^{2}$ & $\pm 8.55\times 10^1$ \\
      			 & Single Element Error  & $\pm 2.09\times 10^{-1} $ & $\pm 2.69$ & $\pm 2.92$ & $\pm 5.65\times 10^{-1}$ & $\pm 2.06 \times 10^{2} $ & $\pm 1.16 \times 10^{2}$ \\
      \midrule
      Moderate & Fiducial Value & 0.027 & 10.5 & 0.8 & 0.1 & 70 & 5 \\
      \cmidrule(l){2-8}
			 & Interferometer Bias & $-5.90\times 10^{-3}$ & $6.39\times 10^{-3}$ & $-7.33\times 10^{-2}$ & $1.50\times10^{-4}$ & $1.77\times 10^{-3}$ & $-1.9\times 10^{-3}$ \\
			 & Single Element Bias & $-9.53 \times 10^{-4}$ & $1.54\times 10^{-2}$ & $-2.63\times 10^{-2}$ & $1.15\times10^{-3}$ & $5.95\times 10^{-2}$ & $-2.36\times 10^{-2}$ \\	
      			 & Interferometer Error & $\pm 4.06\times 10^{-3}$ & $\pm 1.31\times 10^{-2}$ & $\pm 5.86\times 10^{-2}$ & $\pm 2.44 \times 10^{-3}$ & $\pm 3.38\times 10^{-2}$ & $\pm 5.19\times 10^{-2}$ \\
      			 & Single Element Error  & $\pm 4.35 \times 10^{-3}$ & $\pm 3.28 \times 10^{-2}$ & $\pm 7.81\times 10^{-2}$ & $\pm 1.21\times 10^{-2}$ & $\pm 1.74\times 10^{-1}$ & $\pm 2.63\times 10^{-1}$ \\
      \midrule
      Optimistic & Fiducial Value & 0.027 & 8 & 0.5 & 0.2 & 80 & 5 \\
      \cmidrule(l){2-8}
			 & Interferometer Bias & $1.95\times10^{-4}$ & $4.10\times10^{-3}$ & $1.83\times 10^{-3}$ & $-8.62\times 10^{-5}$ &$ -3.80\times 10^{-4}$ & $8.63\times 10^{-4}$ \\
			 & Single Element Bias & $1.26\times10^{-3}$ & $-2.50\times10^{-3}$ & $1.17 \times 10^{-2}$ & $-6.50\times10^{-4}$ & $-1.56 \times 10^{-2}$ & $6.42\times10^{-3}$ \\
      			 & Interferometer Error & $\pm 1.10 \times 10^{-3}$ & $\pm 5.88 \times 10^{-3}$ & $\pm 1.58\times 10^{-2}$ & $\pm 1.65\times 10^{-3}$ & $\pm 1.04\times 10^{-2}$ & $\pm 1.99\times 10^{-2}$ \\
      			 & Single Element Error  & $\pm 1.21 \times 10^{-3}$ & $\pm 6.82 \times 10^{-3}$ & $\pm 1.89 \times 10^{-2}$ & $\pm 6.78\times 10^{-3}$ & $\pm 3.3 \times 10^{-2}$ & $\pm 8.36 \times 10^{-2}$ \\
      \bottomrule
   \end{tabular}
   \caption{\acl{Need to check that I transcribed the new numbers correctly} Parameters, biases, and $1\sigma$ error bars for different scenarios and experiments.}
   \label{tab:params}
\end{table*}

\begin{figure}[h]
	\centering
	\includegraphics[width=0.5\textwidth] {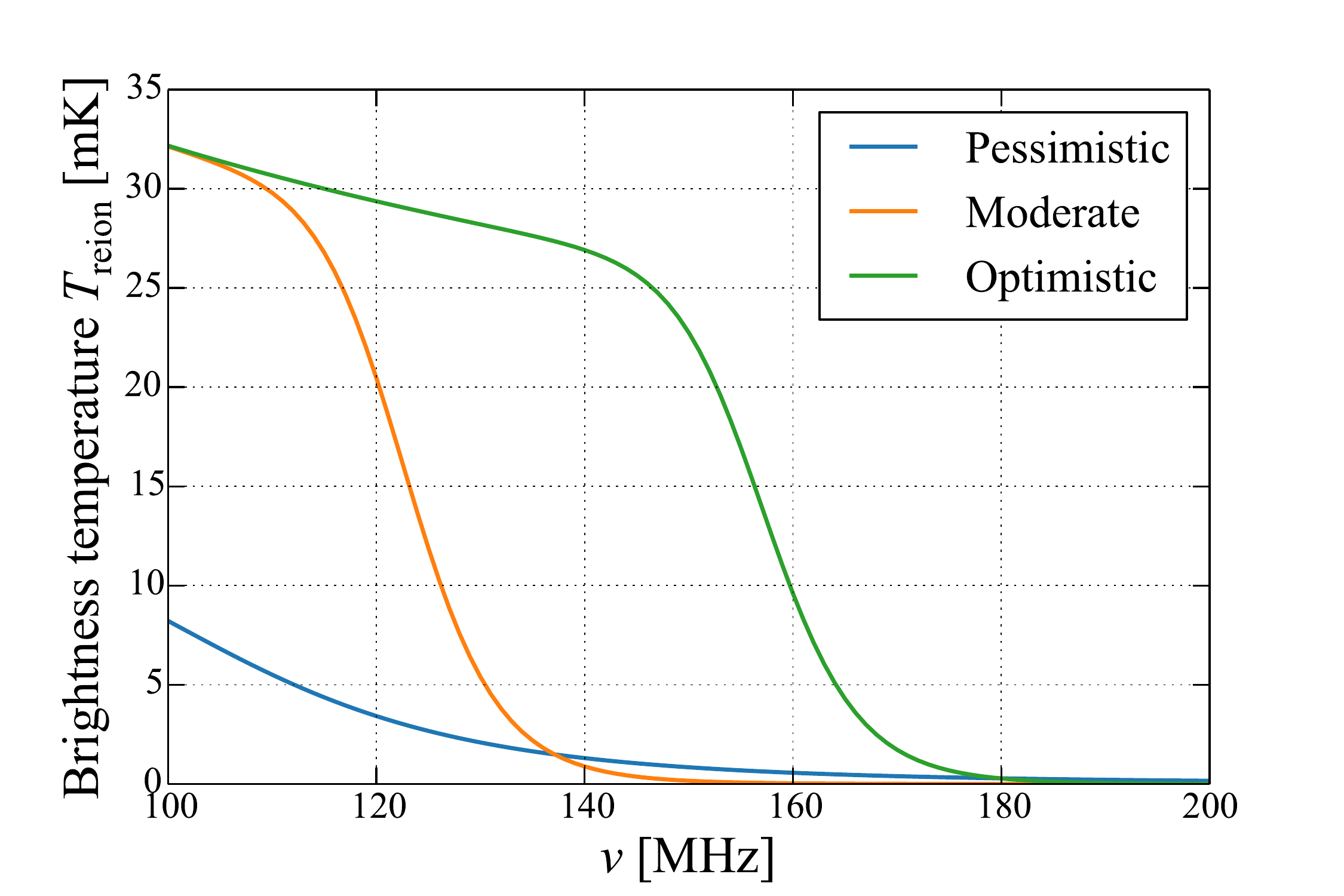}
	\caption{Fiducial model of the brightness temperature during reionization as a function of frequency, following three scenarios: Pessimistic (blue), Moderate (orange), and Optimistic (green). Increasing optimism corresponds to reionization occurring more rapidly and at lower redshifts, making detection easier.}
	\label{fig:reionScenarios}
\end{figure}

\begin{figure}[h]
	\centering
	\includegraphics[width=0.5\textwidth] {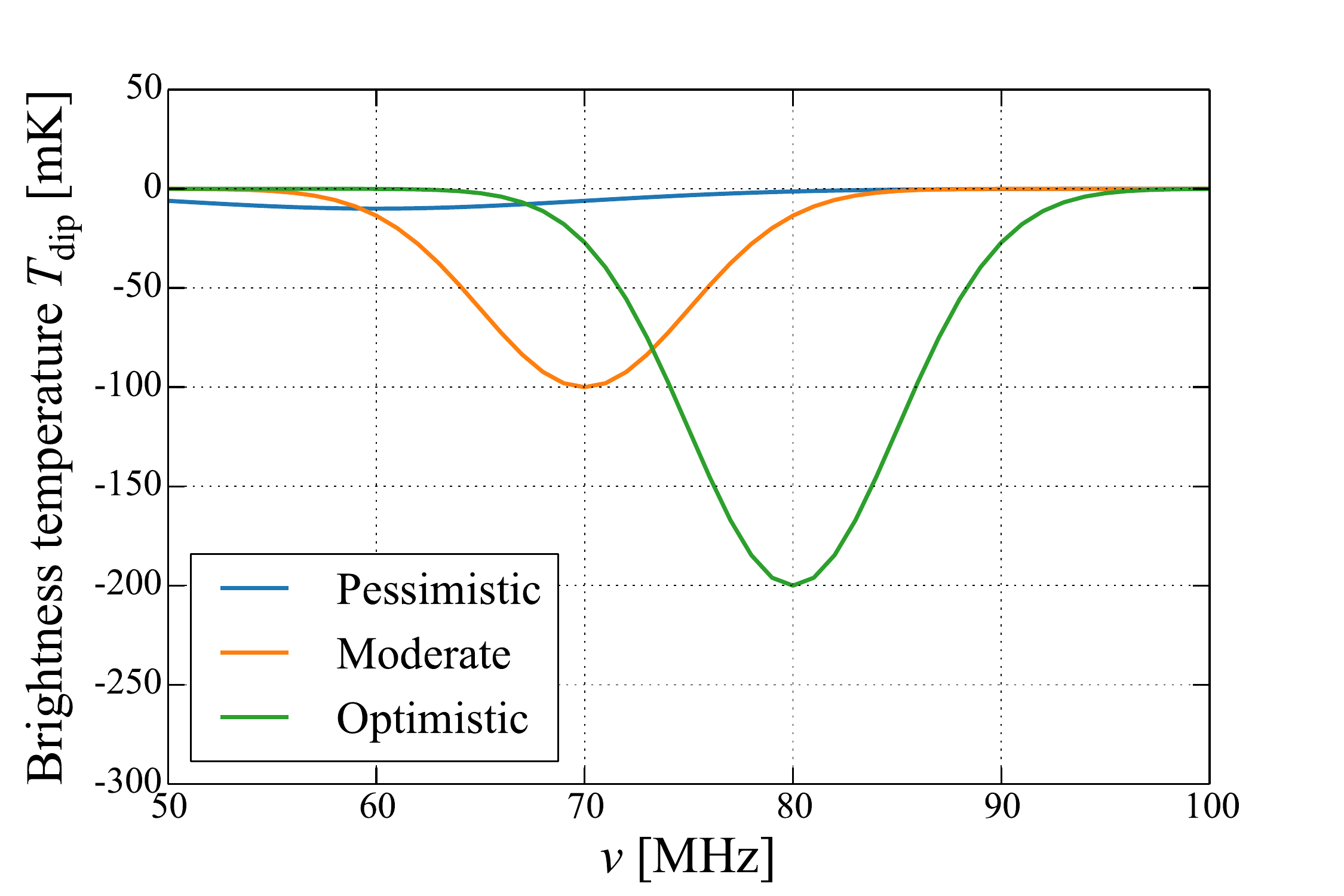}
	\caption{Fiducial model of the brightness temperature during the pre-reionization dip as a function of frequency, following three scenarios: Pessimistic (blue), Moderate (orange), and Optimistic (green). Increasing optimism corresponds to a deeper dip occurring at lower redshifts, making detection easier.}
	\label{fig:dipScenarios}
\end{figure}

The errors on final model parameters will depend on the fiducial ``true" values that are used in our simulations. We consider three different reionization scenarios:
\begin{itemize}
\item Pessimistic reionization scenario, with $(T_{21}, z_r, \Delta z) = (10\,\textrm{mK}, 12, 3) $. With reionization occurring in a rather extended fashion at relatively high redshifts, this scenario should be the most difficult one to detect, since foregrounds are brighter at high redshifts. Additionally, an extended reionization scenario more closely mimics smooth foregrounds.
\item Moderate reionization scenario, with $(T_{21}, z_r, \Delta z) = (27\,\textrm{mK}, 10.5, 0.8) $. This model is motivated by the best-fit value of $z_r$ from the Wilkinson Microwave Anisotropy Probe \citep{WMAP9}. The value of $T_{21}$ is taken from theoretical calculations \citep{PritchardLoeb2010}, while $\Delta z$ is chosen to be neither too extended nor too abrupt.
\item Optimistic reionization scenario, with $(T_{21}, z_r, \Delta z) = (27\,\textrm{mK}, 8, 0.5) $. This scenario is motivated by recent optical and infrared observations \citep{fan_et_al2006,bolton_et_al2011,treu_et_al2013,Faisst_et_al2014}. Having said this, our choice of $(T_{21}, z_r, \Delta z) = (27\,\textrm{mK}, 8, 0.5)$ is not taken from any of these optical/infrared studies in particular, since there is considerable uncertainty in how the observations should be interpreted in the context of reionization. Rather, we simply wish to roughly capture the fact that generically, such observations favor reasonably rapid reionization at lower redshifts, in contrast to what is suggested by CMB experiments.
\end{itemize}
Since measurements of the Gunn-Peterson trough strongly suggest that reionization is complete by $z \sim 6$ \citep{fan_et_al2006}, we choose to impose this as a prior in our Fisher matrix projections. We will find that only the pessimistic scenario is affected by the prior in a non-negligible way. We therefore implement our prior by assuming that $z_r$ and $\Delta z$ are both already known to within $\pm 3$ at $1\sigma$, such that a $2\sigma$ fluctuation would be required for the IGM to be substantially neutral at $z \sim 6$ in the pessimistic scenario.

With the pre-reionization dip we again consider three scenarios of varying degrees of optimism, although without the CMB as a guide, the parameters here are somewhat more arbitrary:
\begin{itemize}
\item Pessimistic pre-reionization scenario, with $(A, \nu_0, \Delta z) = (10\,\textrm{mK}, 60\,\textrm{MHz}, 10\,\textrm{MHz}) $.
\item Moderate pre-reionization scenario, with $(A, \nu_0 \Delta z) = (100\,\textrm{mK}, 70\,\textrm{MHz}, 5\,\textrm{MHz}) $.
\item Optimistic pre-reionization scenario, with $(A, \nu_0, \Delta z) = (200\,\textrm{mK}, 80\,\textrm{MHz}, 5\,\textrm{MHz}) $.
\end{itemize}

\begin{figure*}[t]
	\centering
	\includegraphics[width=1.00\textwidth,trim=3cm 2cm 3cm 0cm,clip] {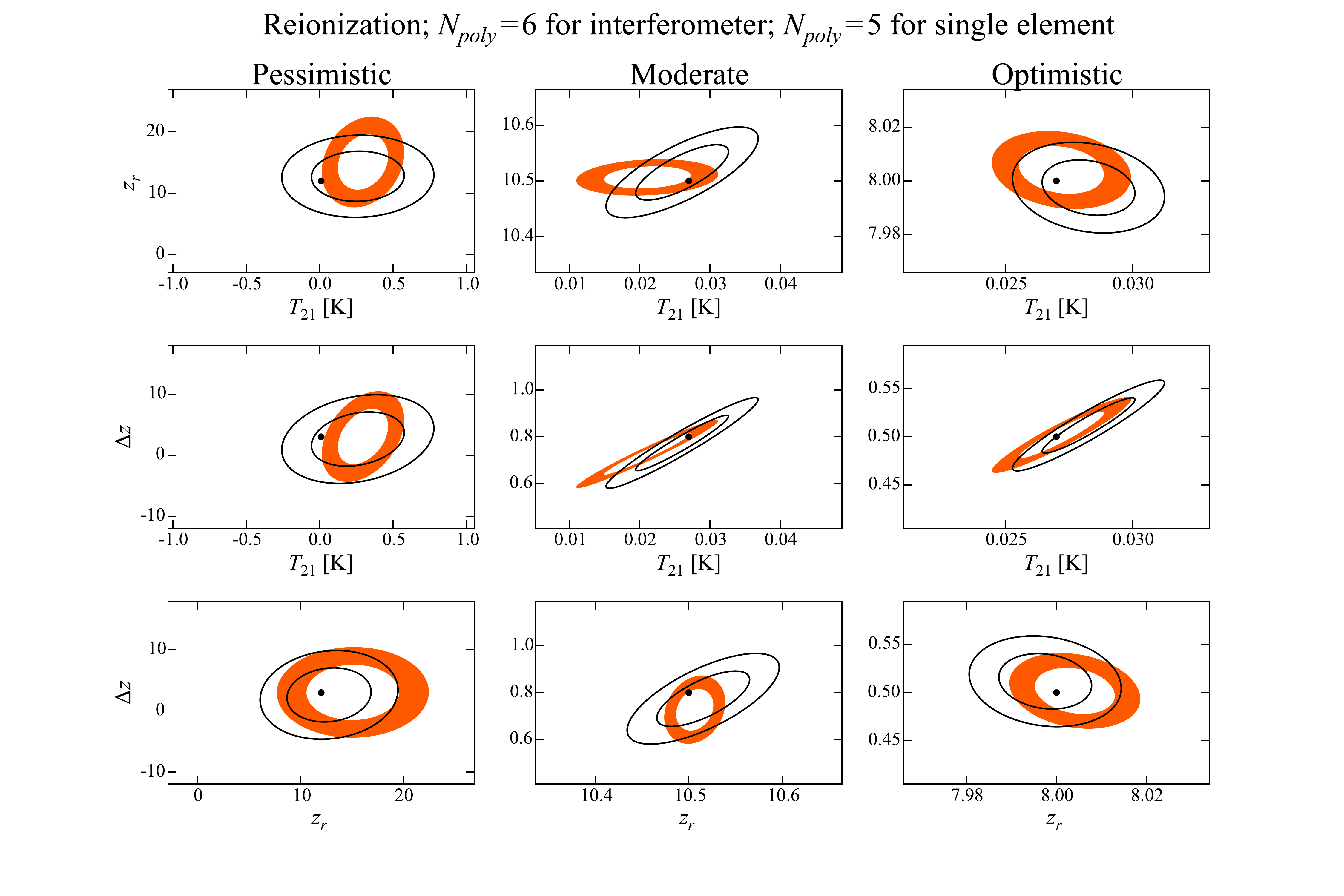}
	\caption{\acl{Updated this} Pairwise parameter contours for interferometer and single element experiments probing reionization. For the interferometer, the orange contours correspond to the $95\%$ confidence regions, and the white areas within the orange contours correspond to the $68\%$ confidence region. For the single-element experiment, the outer solid black contour line bounds the $95\%$ confidence region, while the inner black contour bounds the $68\%$ confidence region. The black dot locates the fiducial value to be recovered and thus indicates the bias in each detection. The interferometer experiment used an $6\times 6$ array of antennas with a primary beam FWHM of $38.7^\circ$ at $100\,\textrm{MHz}$. Both experiments had an integration time of 300 hours. For the interferometric measurement, a $6$th order logarithmic polynomial was needed to model the foreground spectra sufficiently well, whereas with the single element experiment only a $5$th order polynomial was required.}
	\label{fig:reionContoursPoly7Poly7}
\end{figure*}

\begin{figure*}[t]
	\centering
	\includegraphics[width=1.00\textwidth,trim=3cm 2cm 3cm 0cm,clip] {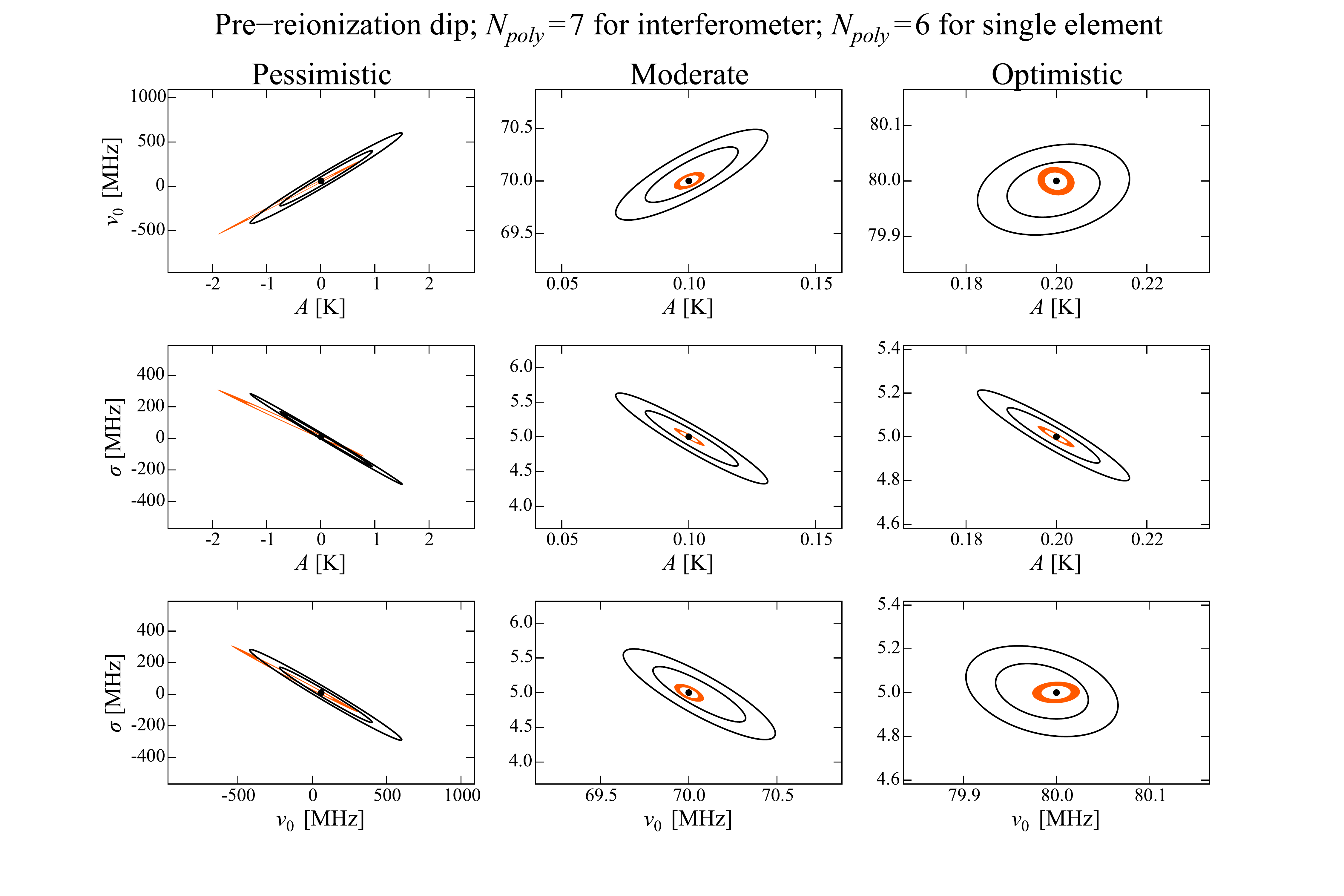}
	\caption{\acl{Updated the figure and the caption} Same as Figure \ref{fig:reionContoursPoly7Poly7}, but for experiments targeting the pre-reionization dip. Primary beams are set to have a FWHM of $38.7^\circ$ at $50\,\textrm{MHz}$. The errors from the interferometric set-up are slightly smaller, thanks to our choice of a slightly larger-than-necessary array as our fiducial model.}
	\label{fig:dipContoursPoly7Poly6}
\end{figure*}

These scenarios are depicted in Figures \ref{fig:reionScenarios} and  \ref{fig:dipScenarios}. The corresponding model parameters are summarized in Table \ref{tab:params}, along with the associated Fisher matrix projections for each parameter's bias and $1\sigma$ error bar after having marginalized over the other parameters. Pairwise parameter contours are shown in Figures \ref{fig:reionContoursPoly7Poly7} and Figures \ref{fig:dipContoursPoly7Poly6}. The results for interferometers are shown using the filled regions, with orange portions signifying $95\%$ confidence regions and the enclosed white portions signifying $68\%$ confidence regions. The black contours demarcate the $68\%$ and $95\%$ confidence regions for the single-element experiments.

\acl{Updated this paragraph to reflect new N-poly}
Under our insistence that the parameter biases from residual foregrounds be subdominant compared to the parameter errors, we find that for reionization, it is necessary to set $N_\textrm{poly} = 6$ for the interferometer whereas for the single-element experiment $N_\textrm{poly} = 5$ suffices. With the pre-reionization dip, we require $N_\textrm{poly} = 7$ for the interferometer and $N_\textrm{poly} = 6$ for the single element. The general trend of higher $N_\textrm{poly}$ for the interferometer reflects the inherently more chromatic nature of interferometry. The generally higher $N_\textrm{poly}$ needed to detect the pre-reionization dip (compared to that needed for to detect reionization) reflects the stronger foregrounds at lower frequencies.

Immediately obvious from Figures \ref{fig:reionScenarios} and \ref{fig:dipScenarios} is the fact that the pessimistic scenarios will be extremely difficult to measure, whether using an interferometer or a single element. With those scenarios, the cosmological signals are simply too extended, and occur at too high redshifts for them to be easily distinguished from the bright foregrounds. Indeed, one can see from Table \ref{tab:params} that the errors on $z_r$ and $\Delta z$ are all approximately $\pm 3$, indicating that constraints are driven entirely by prior information.

\acl{softened claims a little} Encouragingly, we see that both the moderate and the optimistic scenarios should be detectable by both types of instrument. Importantly, one sees that the interferometer performs just as well as the single-element experiment does. For the pre-reionization dip, we often even get slightly smaller error bars with the interferometer. This is because we conservatively chose to simulate a $6\times 6$ array, when a smaller---and therefore higher-noise---array would have sufficed according to our calculations in Section \ref{sec:numElems}. This translates into tighter constraints in the moderate and optimistic scenarios, the errors turn out to be mostly thermal noise dominated. We note that this is not universally the case, and happens only when the cosmological signals in question are sufficiently different from the foregrounds (hence the fact that our discussion here applies mostly to the pre-reionization dip instead of reionization itself). To understand this, consider the effect of varying $N_\textrm{poly}$ in one's analysis pipeline. With $N_\textrm{poly}$ set too low, the bias is too large and the measured parameters are inaccurate; with $N_\textrm{poly}$ set too high, the degeneracies between the foreground model and the cosmological model result in a large variance and the parameters are imprecise. If the cosmological signal is different enough from the foregrounds (as is the case in the moderate and optimistic scenarios), there are a set of intermediate $N_\textrm{poly}$ where the bias is small but the degeneracies have yet to dominate, resulting in a thermal-noise dominated measurement. This is particularly desirable because it means that measurements with more integration time will yield better constraints. \mep{fixed sentence per reviewer comment} On the other hand, the cosmological signals in the pessimistic cases are sufficiently similar to the foregrounds that the variances increase rather quickly with increasing $N_\textrm{poly}$, and come to dominate before the biases become negligible. There is thus never a thermal noise dominated regime, and indeed, we find that increasing the integration time does very little to improve the constraints in the pessimistic scenario.

In summary, the results here suggest that an interferometric measurement of the global signal may be an interesting, viable alternative to single-element experiments. Given the small number of antennas necessary for a competitive interferometric array, one could even imagine collecting both auto-correlation and cross-correlation (visibility) data between antennas, analyzing data from the two modes separately as a way to cross-check the final results.

\section{Conclusions}
\label{sec:Conc}

In this paper, we explored the unusual concept of measuring the global $21\,\textrm{cm}$ signal with an interferometer. We established general design principles for a global signal interferometer as well as a general framework for data analysis. Numerical forecasts confirmed the viability of an interferometric measurement of the global signal, with our fiducial interferometer performing comparably to conventional single-element experiments.

Balancing sensitivity requirements and foreground mitigation considerations, we found that an optimal array design consists of a small grid of closely packed antennas, each with a FWHM beam size of $\sim40^\circ$ at the lowest frequencies. We chose a two-step process for our general analysis method: first we estimated the spatial monopole at each frequency channel; then we combined the estimates into a single global signal spectra and performed a final foreground subtraction. During the first step, we found that overly aggressive downweightings of angular foreground modes led to spectral features that compromised our ability to remove foregrounds. Based on this, we recommend a ``simple is best" approach, where one essentially performs a linear fit for the best-fit monopole given a set of visibilities weighted by their sensitivity to the $\mathbf{u}=0$ mode. In the second step, we found that fitting the single-element-derived spectra to $6$th order logarithmic polynomials were sufficient to reduces foreground residuals to acceptable levels, whereas the interferometer-derived spectra required $7$th order logarithmic polynomials. However, the final parameter errors are comparable between the two experiment types, and give tight constraints on the pre-reionization era and reionization itself under reasonably non-pessimistic scenarios.

\acl{Added line about antenna design} Future work must address a number of systematic issues that will inevitably arise with an experiment in the manner described in this paper. From our discussion in Section \ref{sec:PictorialToyIntro}, it is clear that an interferometer's response to the monopole will depend sensitively on antenna design. Our recommendation of a closely packed array configuration will also likely require careful engineering attention to reduce the possibility of mutual coupling between adjacent antenna elements. Additionally, ionospheric fluctuations can cause systematics that do not integrate down with time \citep{DattaIonosphere}. Luckily, these systematics are reasonably spectrally smooth and thus one may be hopeful that techniques can be developed to mitigate them. In any case, we have shown that an interferometric measurement of the global signal, while unusual, has the potential to rival those from single-element experiments. Importantly, certain classes of systematics, such as thermal noise bias, that are absent from interferometers. (Of course, thermal noise \emph{variance} is present in both types of experiment). Given the technical challenges of $21\,\textrm{cm}$ cosmology, one should thus explore as many complementary experimental approaches as possible. Doing so will maximize the chances of a near-term detection of the cosmological $21\,\textrm{cm}$ signal, providing a crucial first step towards an exquisite understanding of an excitingly unexplored portion of our cosmic timeline.

\section{Acknowledgements}
It brings the authors great pleasure to thank Zaki Ali, John Carlstrom, Carina Cheng, Casey Law, Dan Marrone, Tim Pearson, Dick Plambeck, Jonathan Pritchard, Ravi Subrahmanyan, Eric Switzer, Nithyanandan Thyagarajan, and Harish Vedantham for useful discussions. This research used resources of the National Energy Research
Scientific Computing Center, a DOE Office of Science User Facility 
supported by the Office of Science of the U.S. Department of Energy 
under Contract No. DE-AC02-05CH11231. This work was supported in part by the NSF CAREER award No. 1352519 and the NSF AST grant No. 1129258.

\pagebreak
\appendix
\section{Proof of minimum-variance property of global signal estimator}
\label{minVarProof}

In this Appendix, we provide a constructive proof that choosing $\M$ to be diagonal (as we do in this paper starting in Section \ref{sec:BetterM}) minimizes the variance of our spherical harmonic mode estimator $\xhat$, provided $\Hmat$ is selected to be $\mathbf{N}^{-1}$ in Equation \eqref{eqn:xhat}. Given this choice, Equation \eqref{eq:NoiseMatrixSigma} for the covariance $\boldsymbol \Sigma$ of $\xhat$ reduces to  $\M \mathbf{B} \M^\dagger$, where $\mathbf{B} \equiv \Q^\dagger \N^{-1} \Q$. With this notation, the window function matrix becomes $\W = \M \mathbf{B}$.

To derive a minimum-variance estimator, we minimize the diagonal elements of $\boldsymbol \Sigma$ subject to the constraint that the window functions satisfy $\W_{ii} = (\M \mathbf{B})_{ii} =1$. Introducing a Lagrange multiplier $\lambda_i$, we seek to minimize the quantity
\begin{equation}
L = (\M \mathbf{B} \M^\dagger)_{ii} - \lambda_i  (\M \mathbf{B})_{ii}.
\end{equation}
Differentiating with respect to the real and imaginary parts of a generic component $\M_{jk}$ of $\M$, one obtains
\begin{equation}
\frac{\partial L}{\partial \textrm{Re} \M_{jk}} = (\mathbf{B} \M^\dagger + \M \mathbf{B})_{jk} - \lambda_j \mathbf{B}_{kj} =0
\end{equation}
and
\begin{equation}
\frac{\partial L}{\partial \textrm{Im} \M_{jk}} = (\mathbf{B} \M^\dagger - \M \mathbf{B})_{jk} - \lambda_j \mathbf{B}_{kj} =0,
\end{equation}
where we have set each expression to zero in order to perform a minimization. Now, recall that $\mathbf{B}$ is Hermitian. However, the combination $\mathbf{B} \M^\dagger - \M \mathbf{B}$ is by construction anti-Hermitian. For the second equation to hold, then, we require that $\mathbf{B} \M^\dagger = \M \mathbf{B}$, so that the anti-Hermitian portion vanishes identically. With this, the first equation reduces to $2 \M \mathbf{B} = \boldsymbol \Lambda \mathbf{B}$, where $\boldsymbol \Lambda_{ij} = \lambda_i \delta_{ij}$. Acting on both sides with $\mathbf{B}^{-1}$ then gives $\M = \boldsymbol \Lambda / 2$, which tells us that $\M$ must be diagonal. To fix the values of the Lagrange multipliers along the diagonal, we use our constraint $\W_{ii} = (\M \mathbf{B})_{ii} =1$ to obtain $\M_{ij} = \delta_{ij} / \mathbf{B}_{ii}$. Recalling the definition of $\mathbf{B}$, we see that this is precisely the form of Equation \eqref{eq:diagM}, completing our proof.

\section{Global signal Fisher matrix forecasting methods}
\label{fisher}
In this Appendix, we review the Fisher matrix formalism used in \citet{PritchardLoeb2010} and \citet{BernardiLEDA} \mep{Added citation} to forecast the error bars and biases in a global signal measurement. We do not claim any originality here, and include a description of the formalism only for completeness and consistency of notation.

Suppose we group our final estimate of the global signal $\widehat{T}_0 (\nu)$ into a vector $\widehat{\mathbf{T}}$, so that each component corresponds to the measured global signal value at a particular frequency. With this notation, the Fisher information matrix is defined as
\begin{equation}
\F_{ij} \equiv -\left \langle \frac{\partial^2 \ln \mathscr{L}(\boldsymbol \theta | \widehat{\mathbf{T}})}{\partial \theta_i \partial \theta_j} \Big{|}_{\boldsymbol \theta_0} \right \rangle
\end{equation}
where $\mathscr{L}(\boldsymbol \theta | \widehat{\mathbf{T}})$ is the likelihood function of a set of model parameters $\boldsymbol \theta$ given the measurements $\widehat{\mathbf{T}}$, with $\boldsymbol \theta_0$ denoting a set of fiducial values for the parameters. Now, recall that $\widehat{\mathbf{T}}$ contains the measured global signal spectrum prior to the smooth foreground fitting described in Section \ref{sec:fitting}. It therefore contains not only the cosmological signal (parameterized by one of the forms given in Section \ref{sec:SimResults}), but also foregrounds and noise, and may be written as
\begin{equation}
\widehat{\mathbf{T}}= \widehat{\mathbf{T}} _\textrm{cosmo} + \widehat{\mathbf{T}}_\textrm{fg} + \boldsymbol \sigma = \boldsymbol \mu (\boldsymbol \theta) +\delta \widehat{\mathbf{T}}_\textrm{fg} + \boldsymbol \sigma,
\end{equation}
where $\widehat{\mathbf{T}} _\textrm{cosmo}$ is the cosmological signal, $\widehat{\mathbf{T}}_\textrm{fg}$ the foregrounds, and $\boldsymbol \sigma$ the residual instrumental noise. In the second equality, we separated the foreground contribution into a portion accounted for by the smooth spectrum fits described in Section \ref{sec:fitting} and a residual, $\delta \widehat{\mathbf{T}}_\textrm{fg}$. The former contribution is combined with the cosmological signal to give a model $\boldsymbol \mu (\boldsymbol \theta)$ for our spectrum. Since this model contains our foreground fits, the parameter vector $\boldsymbol \theta$ records not only the astrophysical/cosmological parameters such as those in Equations \eqref{eq:Dip} and \eqref{eq:Step}, but also the foreground parameters in Equation \eqref{eq:FgFit}. Assuming that the instrumental noise is Gaussian and has zero mean and covariance $\boldsymbol \Pi \equiv \langle \boldsymbol \sigma \boldsymbol \sigma^t \rangle$, the Fisher matrix takes the form
\begin{equation}
\F_{ij} = \frac{\partial \boldsymbol \mu ^t}{\partial \theta_i} \boldsymbol \Pi ^{-1} \frac{\partial \boldsymbol \mu}{\partial \theta_j}.
\end{equation}
Once the Fisher matrix has been computed, the smallest possible error $\Delta \theta_i$ (in the information theoretic sense) on the parameter $\theta_i$ is obtained by calculating $\Delta \theta_i = \sqrt{(\F^{-1})_{ii}}$. While an actual experiment may not necessarily deliver error bars that are as tight as this, the predictions of the Fisher matrix formalism are nonetheless a useful guide for experimental design.

Prior information can also be incorporated into our parameter estimates. If observations from other probes have already constrained the $i$th parameter to within an error of $\varepsilon_i$, this can be accounted for by adding $\varepsilon_i^{-2}$ to the $i$th diagonal element of the Fisher matrix. The revised parameter errors will be smaller not just for the $i$th parameter, but for all the other parameters as well, since knowing the $i$th parameter better can help to break degeneracies.

So far, we have only concerned ourselves with the variance of the final parameter constraints. However, foreground residuals will cause more than a spread in the parameter fits---they will also bias the fits in a systematic way. This bias is given by
\begin{equation}
\delta \theta_i = \sum_{j} (\F^{-1})_{ij} \frac{\partial \boldsymbol \mu ^t}{\partial \theta_j} \boldsymbol \Pi ^{-1} \delta \widehat{\mathbf{T}}_\textrm{fg}.
\end{equation}
In Section \ref{sec:SimResults}, $\delta \xhat_\textrm{fg}$ is obtained by running Monte Carlo simulations. We generate realizations of the foreground sky, which are then fed through a simulation of a measurement and data analysis. Averaging over simulations, the resulting spectra are then fit to logarithmic polynomials. The residuals (i.e., $\delta \widehat{\mathbf{T}}_\textrm{fg}$) are given by Equation \eqref{eq:FgFit}.

\bibliographystyle{apj}
\bibliography{globalSigInterferometer}{}

\end{document}